\documentclass{PoS}

\usepackage{wrapfig}

\title{Status of Dark Matter Searches (Rapporteur Talk)}

\ShortTitle{Status of Dark Matter Searches}

\author{\speaker{Carsten Rott}\\
        Department of Physics, Sungkyunkwan University, Suwon 440-746, Korea\\
        E-mail: \email{rott@skku.edu}}

\abstract{This article reviews the status of the field of dark matter as of summer 2017, when it was discussed at 35th International Cosmic Ray Conference (ICRC 2017) in Busan, Korea. It is the write-up of a rapporteur talk on the status of dark matter searches given at the conference.}

\FullConference{35th International Cosmic Ray Conference --- ICRC2017\\
		10--20 July, 2017\\
		Bexco, Busan, Korea}

\begin{document}

\section{Introduction}

There is compelling observational evidence for the existence of dark matter~(DM) on all distance scales. Starting from imprints on the cosmic microwave background to large scale structures, galaxy clusters, dwarf spheriodal galaxies and even in our own Milky Way. 
Recent results from the Planck satellite indicate that dark matter accounts for 83\% of the cosmological matter density~\cite{Ade:2015xua}.

Although knowledge of the underlying nature of dark matter remains elusive, a variety of theories provide candidate particles that can explain the indirect evidence~\cite{Bertone:2004pz}. Most attractive have been dark matter models that also solve fundamental problems in particle physics such as the hierarchy problem. Particle dark matter (denoted as $\chi$) is expected to be produced in the early hot and dense universe. Particles are neutral, stable (or long lived), weakly interacting, and might exhibit the properties of a Weakly Interacting Massive Particle (WIMP)~\cite{Steigman:1984ac}.

Dark matter is searched for in three complementary ways, consisting of direct, indirect, and collider searches. Direct searches search for the scattering process of dark matter with ordinary matter, which might be measurable in nuclear recoils. Collider searches aim at recreating the conditions in the early hot universe to produce dark matter particles and study them directly. Evidence of dark matter might present itself in observations of processes that deviate from the Standard Model~(SM) of particle physics predictions.
Indirect dark matter searches look for evidence for dark matter annihilations or decays through the observation of stable messenger particles (such as $\gamma, \nu$, e$^{\pm}$, p / $\bar{\rm p}$) created in the process, see figure~\ref{DM_evidence}. The latter connects closely to the observations of gamma-rays, neutrinos, and charged cosmic rays, which is the main topic at ICRC. 

\begin{wrapfigure}{r}{3.0in}
\begin{center}
\vspace*{-1.0cm}
        \includegraphics[width=0.36\textwidth]{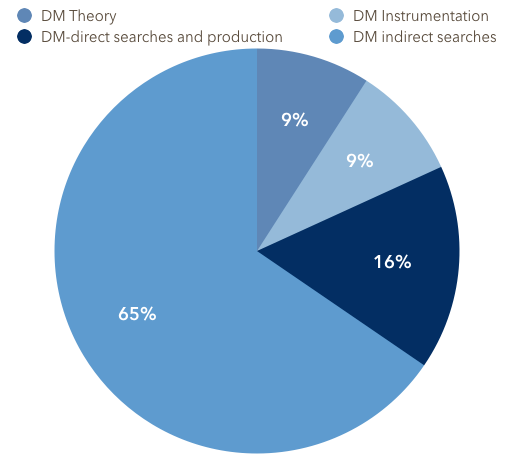}
        \caption{Dark Matter related contributions to the ICRC2017 conference.}
        \label{DM_at_ICRC2017}
\vspace*{-0.6cm}
\end{center}
\end{wrapfigure}        
ICRC brings together experts from direct, indirected, theory dark matter community. While ICRC's main focus is on cosmic ray and astroparticle physics, it also gives important impulses to the dark matter field. (1) Historically many unexplained astrophysical observations have been interpreted as potential dark matter signals. To reliably evaluate the significance of a potential dark matter signal it is essential to understand the uncertainties in the observation. (2) Any unexplained dark matter signal has almost certainly implication on other observation channels. Multi-messenger observations are required and consistent picture has to emerge before any discovery should be claimed. (3) There are many successful examples where instrumentation developed for cosmic ray physics can be used for dark matter searches or existing cosmic-ray data can be utilized to search for dark matter. At ICRC2017 a variety of dark matter related topics were discussed and their relative contributions are summarized in figure~\ref{DM_at_ICRC2017}.

While I try to give a comprehensive review of the dark matter field, I have placed priorities on the dark matter related contributions given at ICRC2017. The reader should both be able to get the general status of the field, while at the same time understand what was presented and discussed at the conference.


\section{Indirect Detection}

\begin{figure}[htb]
        \centering
        \includegraphics[width=0.49\textwidth]{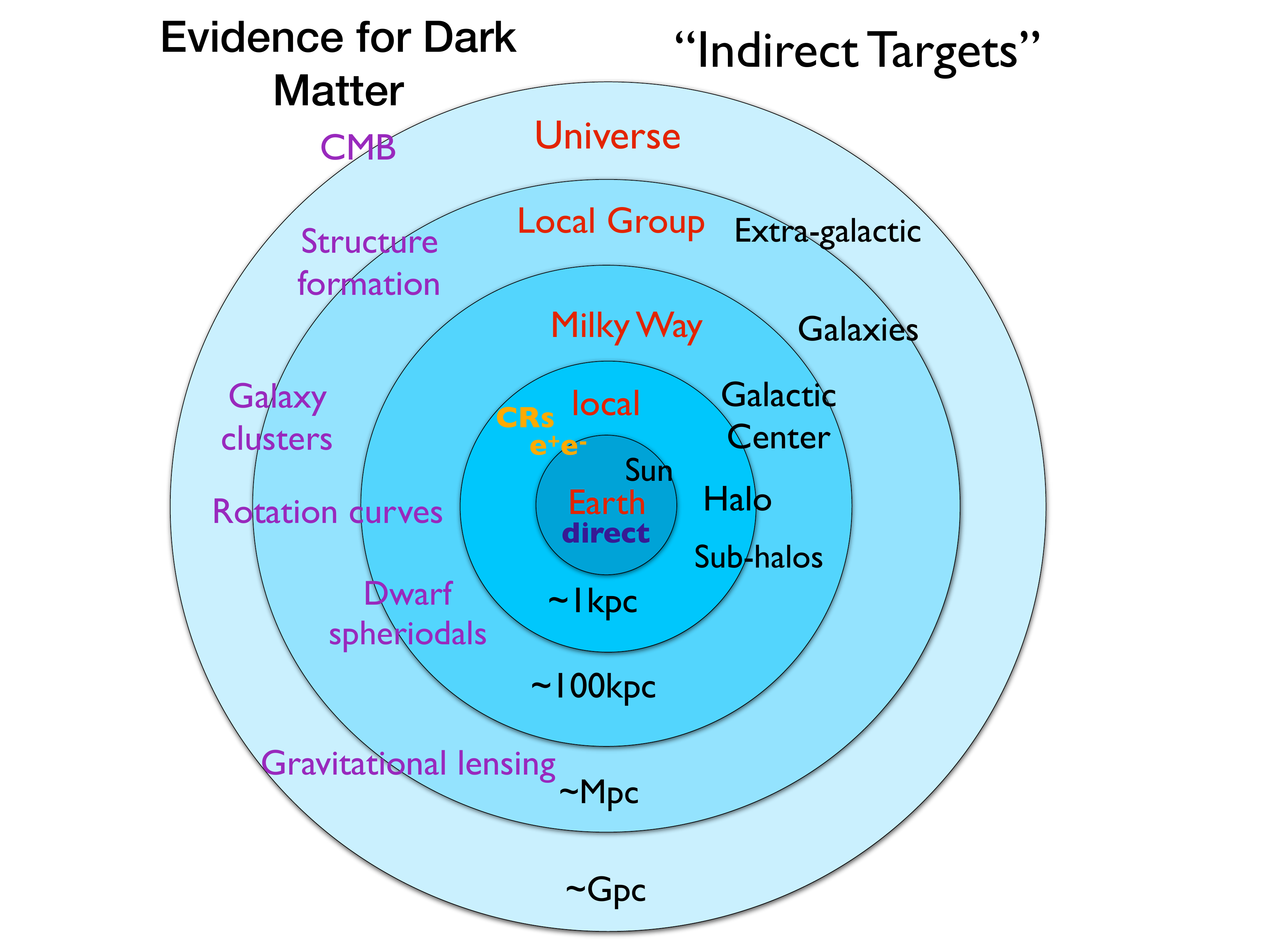}
        \includegraphics[width=0.49\textwidth]{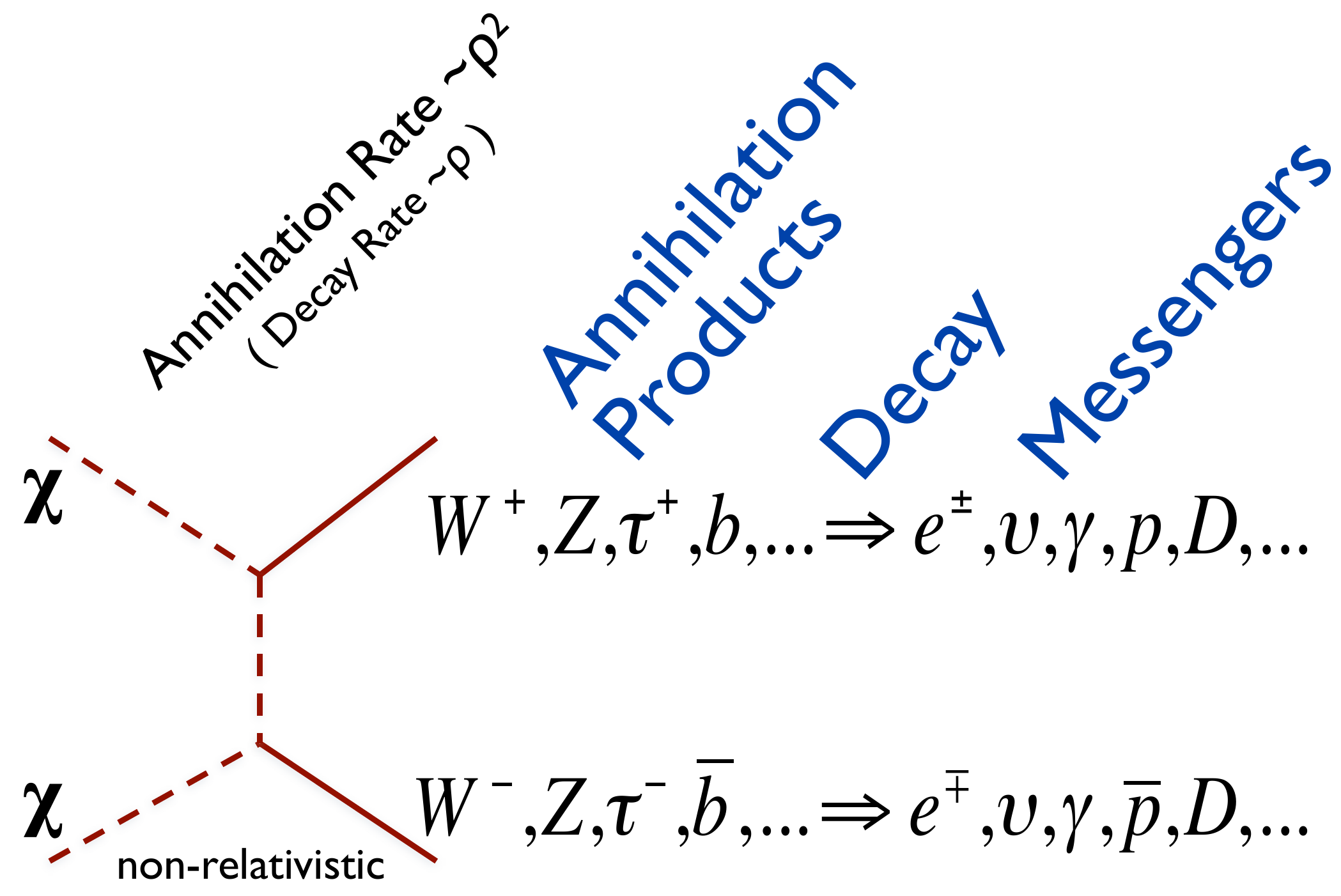}
        \caption{Left: Evidence for dark matter can be found on all scales as well as targets for indirect searches. Right: Principle of indirect dark matter signals.}
        \label{DM_evidence}
\end{figure}

\begin{figure}[htb]
        \centering
        \includegraphics[width=0.79\textwidth]{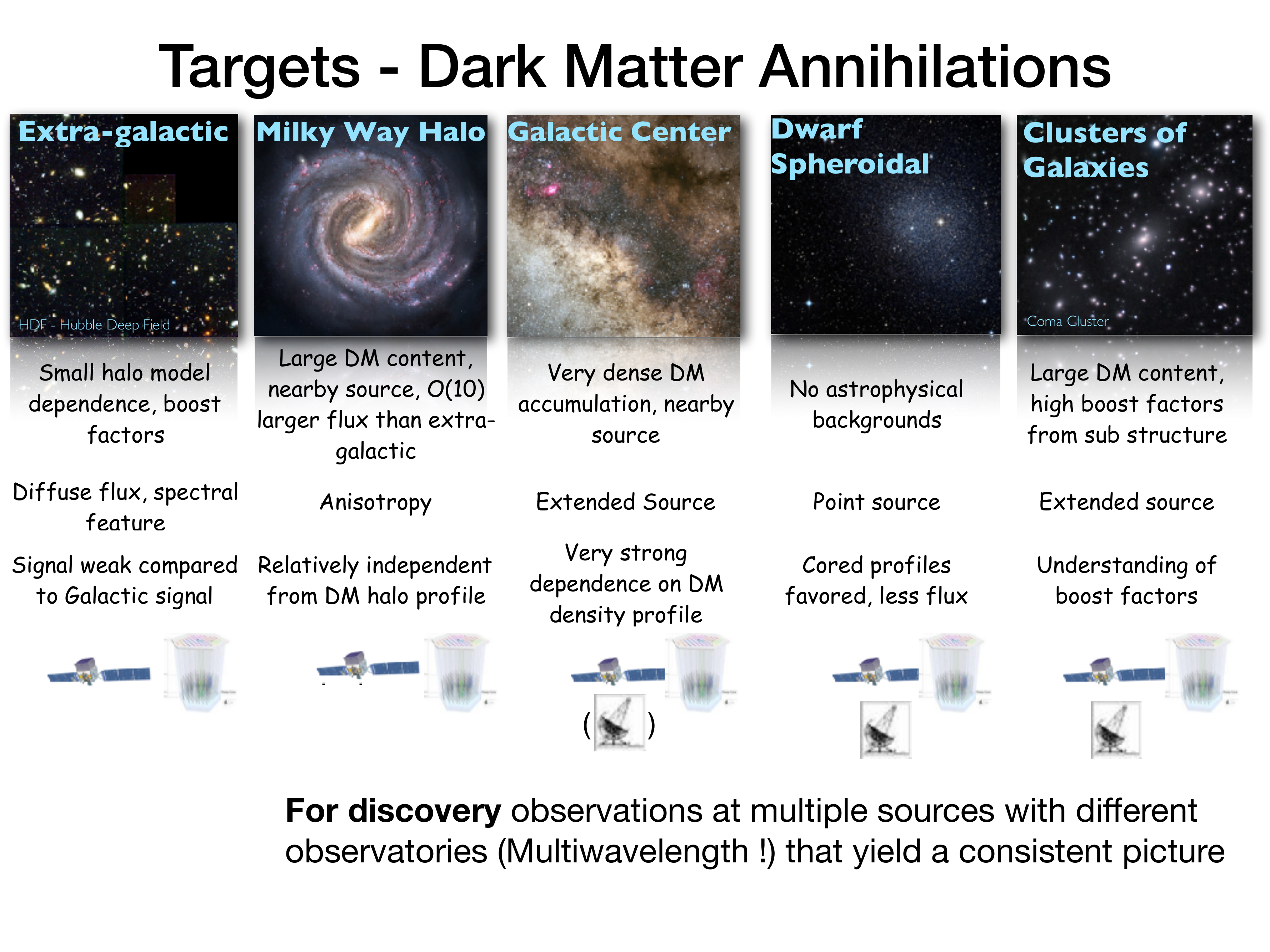}
        \caption{Targets for indirect dark matter searches and challenges and benefits associated with the sources.}
        \label{Targets_indirect}
\end{figure}

Indirect dark matter searches can concentrate on a variety of different targets that each have their own benefits and challenges associated with them as summarized in figure~\ref{DM_evidence} and figure~\ref{Targets_indirect}. Imaging air Cherenkov telescopes (IACTs) are limited by their field of view and the necessity to measure an off-source region to evaluate changing atmospheric backgrounds. Dwarf spheriodal galaxies~(dSph), galaxy clusters, or large galaxies, which all have small source extensions are ideal targets for IACTs. IACTs have large effective areas, however observations are limited to clear dark nights (low duty factor) and internal competition for target selection. Gamma-ray satellites in survey mode observe the entire sky but effective areas are much smaller compared to IACTs. The Galactic center with its dense dark matter concentration is an ideal target however astrophysical backgrounds can fake signals. Neutrino telescope also observe the entire sky making the Galactic halo an ideal target. Event rates however are limited due to the small neutrino interaction cross sections. At TeV~energies neutrino searches become more competitive relative to gamma-ray based searches as the neutrino effective area increases with energy. Charged cosmic rays can also be used to search for dark matter annihilation or decay signals, however uncertainties in the propagation in the Galactic magnetic fields makes a signal prediction rather complicated and difficult to distinguish from other potential sources.   

\subsection{Gamma-rays}

\subsubsection{Searches for dark matter in dwarf spheriodal Galaxies}

Dwarf spheriodal galaxies are one of the best targets to search for self-annihilating dark matter. They combine a high mass to light ratio (dark matter dominated) and typically have no astrophysical backgrounds. Low astrophysical backgrounds can yield robust constraints, however the conversion of flux limits to thermal averaged annihilation cross section, $\langle \sigma_{A} v \rangle$, requires precisely known dark matter distributions (J-factors). dSph are small in extend and hence ideally suited for the observation with IACTs. Intense efforts by IACTs have resulted in stringent bounds on the dark matter self-annihilation cross section as well as the dark matter lifetime, $\tau$. 
Prospects of identifying more dSph make them also exciting targets for all sky observatories that already have existing data. Newly discovered Milky Way dwarf satellites have been studied by Fermi-LAT~\cite{Fermi-LAT:2016uux} and follow numerous searches using dSphs carried out with neutrinos~\cite{Aartsen:2013dxa} and gamma-rays.

\begin{figure}[htb]
        \centering
        \includegraphics[width=0.39\textwidth]{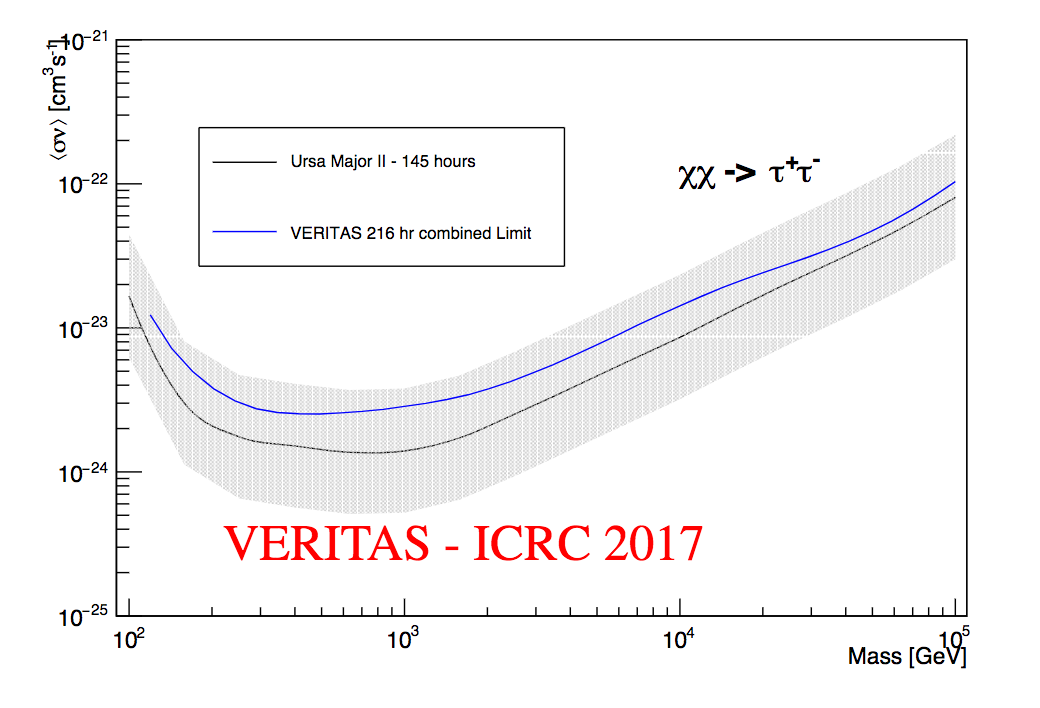}
        \includegraphics[width=0.29\textwidth]{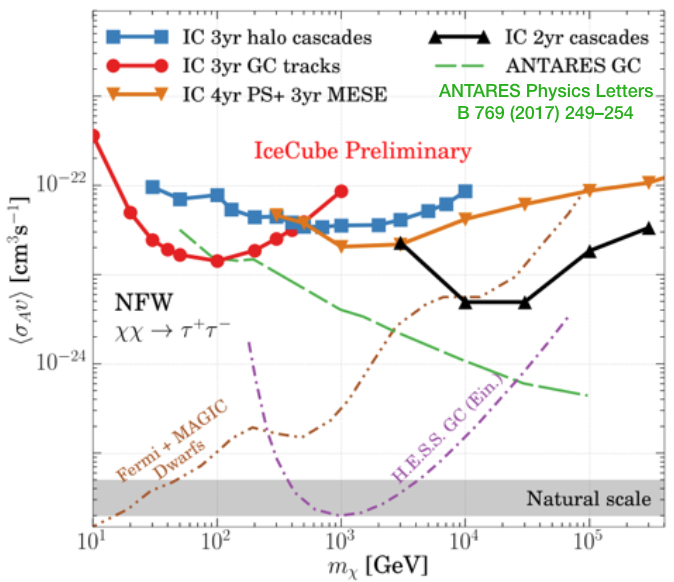}
        \includegraphics[width=0.29\textwidth]{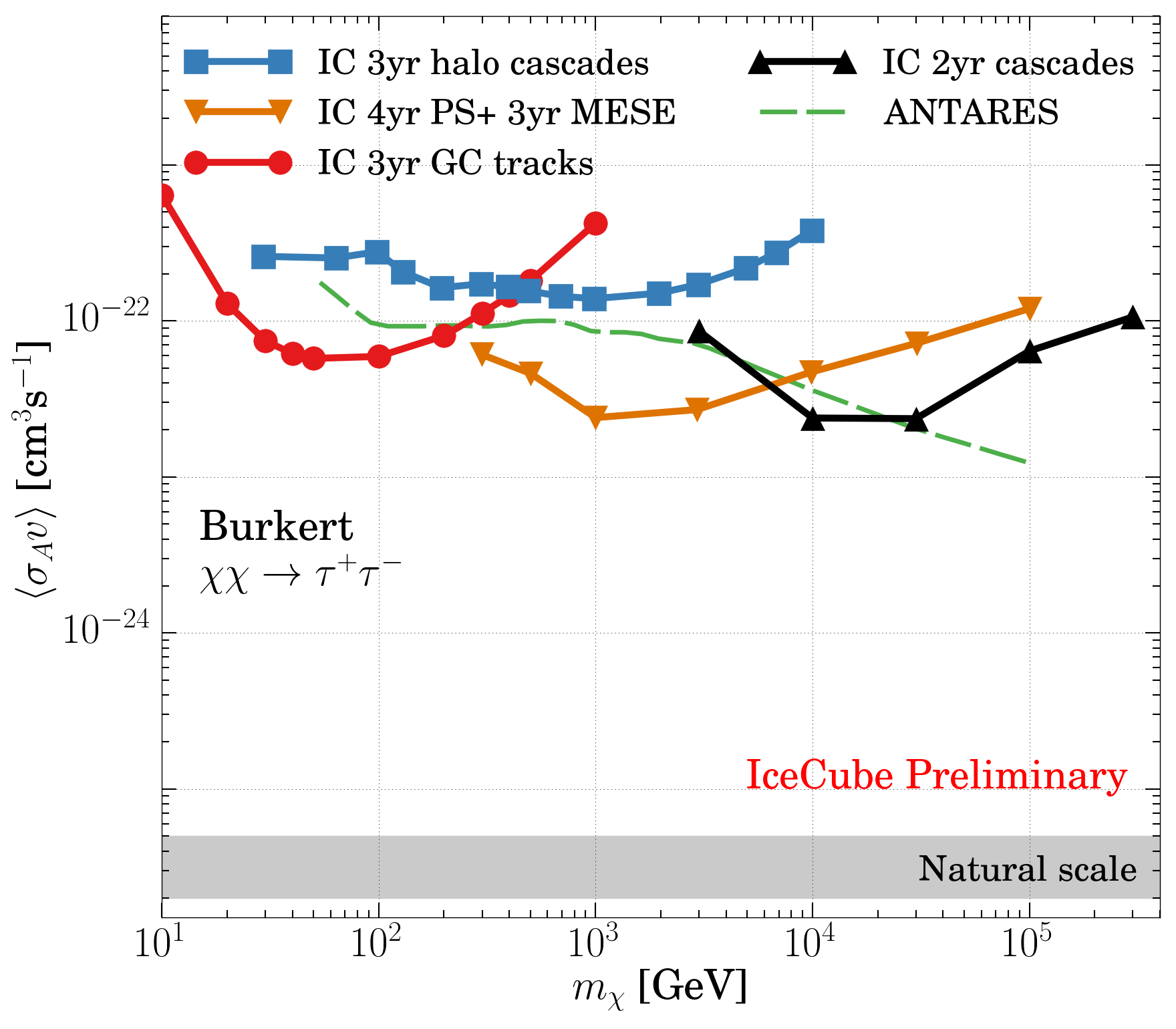}
        \caption{Preliminary limits on the dark matter self-annihilation cross section from VERITAS~\cite{VERITAS_ICRC904} (left) and neutrino telescopes including IceCube~\cite{ICRC_Flis} (middle,right). Dark matter annihilation in $\chi \chi \rightarrow \tau^{+}\tau^{-}$ is assumed. The change in limit based on the assumed dark matter halo profile is shown by comparing the middle (NFW) and right (Burkert) plot (Details see~\cite{VERITAS_ICRC904,ICRC_Flis}). }
        \label{IACT_self_ann}
\end{figure}

The Very Energetic Radiation Imaging Telescope Array System~(VERITAS) obtained limits, shown in figure~\ref{IACT_self_ann}, on the dark matter self-annihilation cross section based on five dSphs observed between 2007 and 2013, using a total of 230 hours (including 92 hours of Segue~1) after data quality selection. VERITAS is an array of four 12~m diameter IACTs located at the Fred Lawrence Whipple Observatory in southern Arizona, USA. VERITAS is sensitive to gamma rays from approximately 85 GeV to greater than 30 TeV with a typical energy resolution of 15\% to 25\% and an angular resolution of $<0.1$~degrees (68\% containment) of 0.1~degrees~\cite{VERITAS_ICRC904} .

\begin{figure}[htb]
        \centering
        \includegraphics[width=0.43\textwidth]{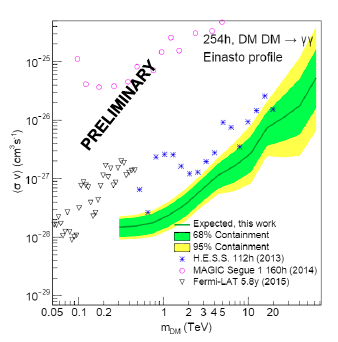}
          \includegraphics[width=0.55\textwidth]{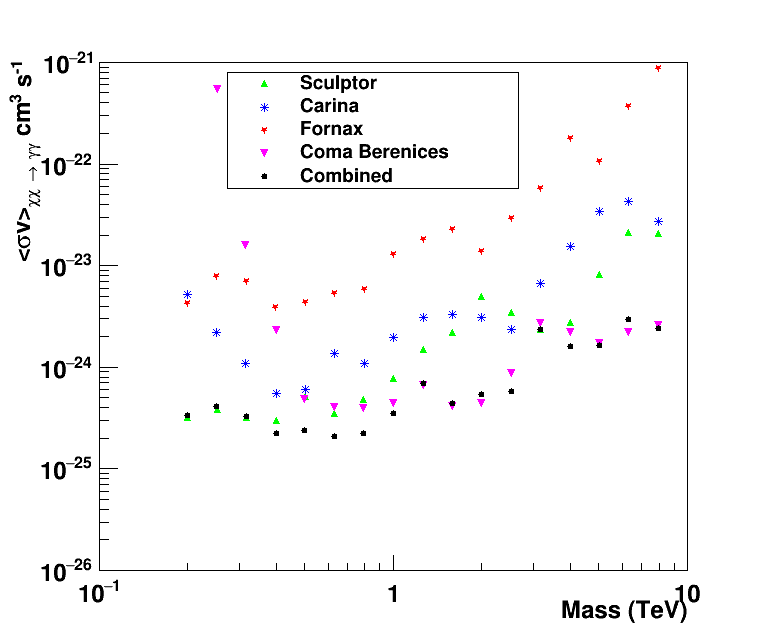}
        \caption{Left: Sensitivity of line search ($\chi \chi \rightarrow \gamma \gamma$) from H.E.S.S. at the Galactic Center. Right: 95\% exclusion limits on the velocity weighted self-annihilation cross section $\langle \sigma_{A} v \rangle$ as a function of the dark matter mass for the combination of four dwarf galaxies bounds from Fornax, Carina, Coma Berenices, and Sculptor~\cite{HESS_893,HESS_905}}
        \label{HESS_line}
\end{figure}

The High Energy Stereoscopic System (H.E.S.S.) is located in the Khomas Highlands, Namibia and consists of four identical imaging atmospheric Cherenkov telescopes and a large fifth telescoped that was added to the center of the array in 2012. Recent results of a search for gamma-ray lines ($\chi \chi \rightarrow \gamma \gamma$) are shown in figure~\ref{HESS_line}.

The High Altitude Water Cherenkov~(HAWC) gamma-ray telescope is located at the Parque Nacional Pico de Orizaba at an altitude of 4100~m. The array consists of 300 water tanks, each of which contains 3$\times$ 8" R5912~PMTs and 1$\times$ 10" R7081-HQE PMT. Tanks have a 7.3~m diameter and a height of 4.5~m. Combined all tanks contain about 55~kT of water and cover an area of $22000~{\rm m}^2$. HAWC is operating in its full detector configuration since March 2015 and partial detector configuration since August 2013.  

One of the great advantages of HAWC over IACTs is that it observes large portions of the sky and is continuously operating. HAWC performed a search for GeV-TeV photons resulting from dark matter annihilation or decay considering dwarf spheroidal galaxies (dSphs), the M31 galaxy and the Virgo cluster as sources with 507~days of data collected in the full detector configuration~\cite{Albert:2017vtb,ICRC_HAWC}. No statistically significant excess from these sources was observed. 95\% confidence level upper limits on the annihilation cross-section and lower limits on the lifetime of dark matter with masses above 1~TeV were computed for various channels. A total of 15~dSphs were considered based on favored declination angles for the HAWC observatory and well understood dark matter content.  The sample included Bootes~I, Canes Venatici~I, Canes Venatici~II, Coma Berenices, Draco, Hercules, Leo~I, Leo~II, Leo~IV, Segue~1, Sextans, Ursa Major~I, Ursa Major~II, Ursa Minor and Triangulum~II. Results are shown in figure~\ref{DM_decay}


Dark Matter Particle Explorer (DAMPE)~\cite{TheDAMPE:2017dtc} was launched on December 17, 2015 from the Jiuquan Satellite Launch Center in the Gobi desert. 
DAMPE consists of a Silicon-Tungsten tracKer-converter (STK) to measures the charges and the trajectories of charged particles, and reconstruct gamma-rays after pair-conversion. It has a Plastic Scintillator strip Detector (PSD) to measures the charge of incident particles and to act as an anti-coincidence detector for gamma ray identification. A hodoscopic BGO (Bismuth Germanium Oxide) imaging  calorimeter and a NeUtron Detector (NUD). DAMPE has a gamma-ray effective area of $0.11~m^2$ and is in a Sun synchronous orbit. 

\begin{figure}[htb]
        \centering
         \includegraphics[angle=90,width=0.89\textwidth]{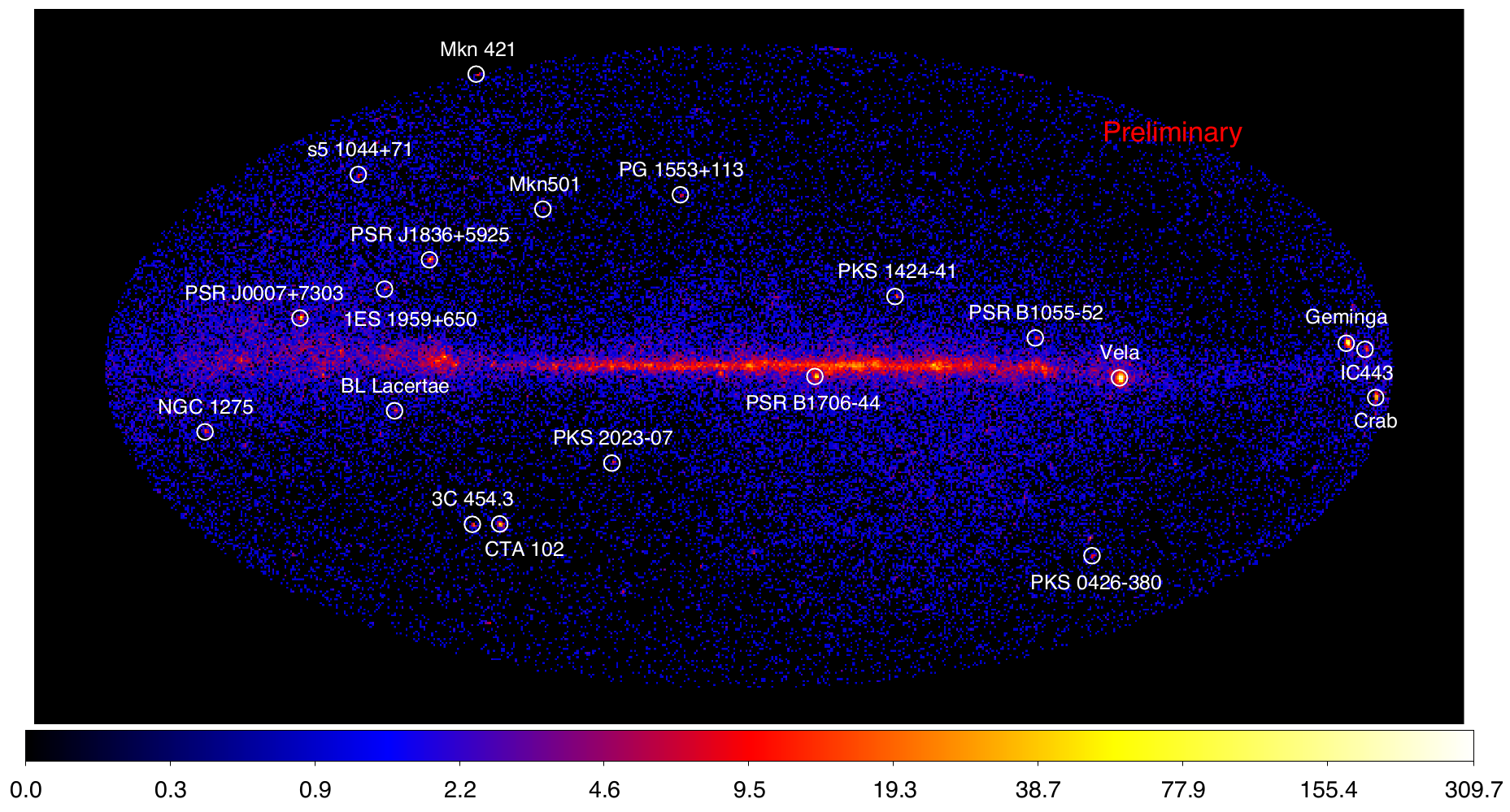}         
        \caption{Preliminary DAMPE skymap using 510~days of data and gamma-rays with reconstructed energies above 2~GeV~\cite{HL_Jin_Chang}}
        \label{DAMPE_skymap}
\end{figure}

At ICRC the DAMPE collaboration showed that the instrument is performing well in orbit~\cite{DAMPE} and showed first results including a preliminary skymap using 510~days of data and gamma-rays with reconstructed energies above 2~GeV~\cite{HL_Jin_Chang}  (see figure~\ref{DAMPE_skymap}).

\subsubsection{Future Gamma-ray Searches}

\begin{figure}[htb]
        \centering
        \includegraphics[width=0.89\textwidth]{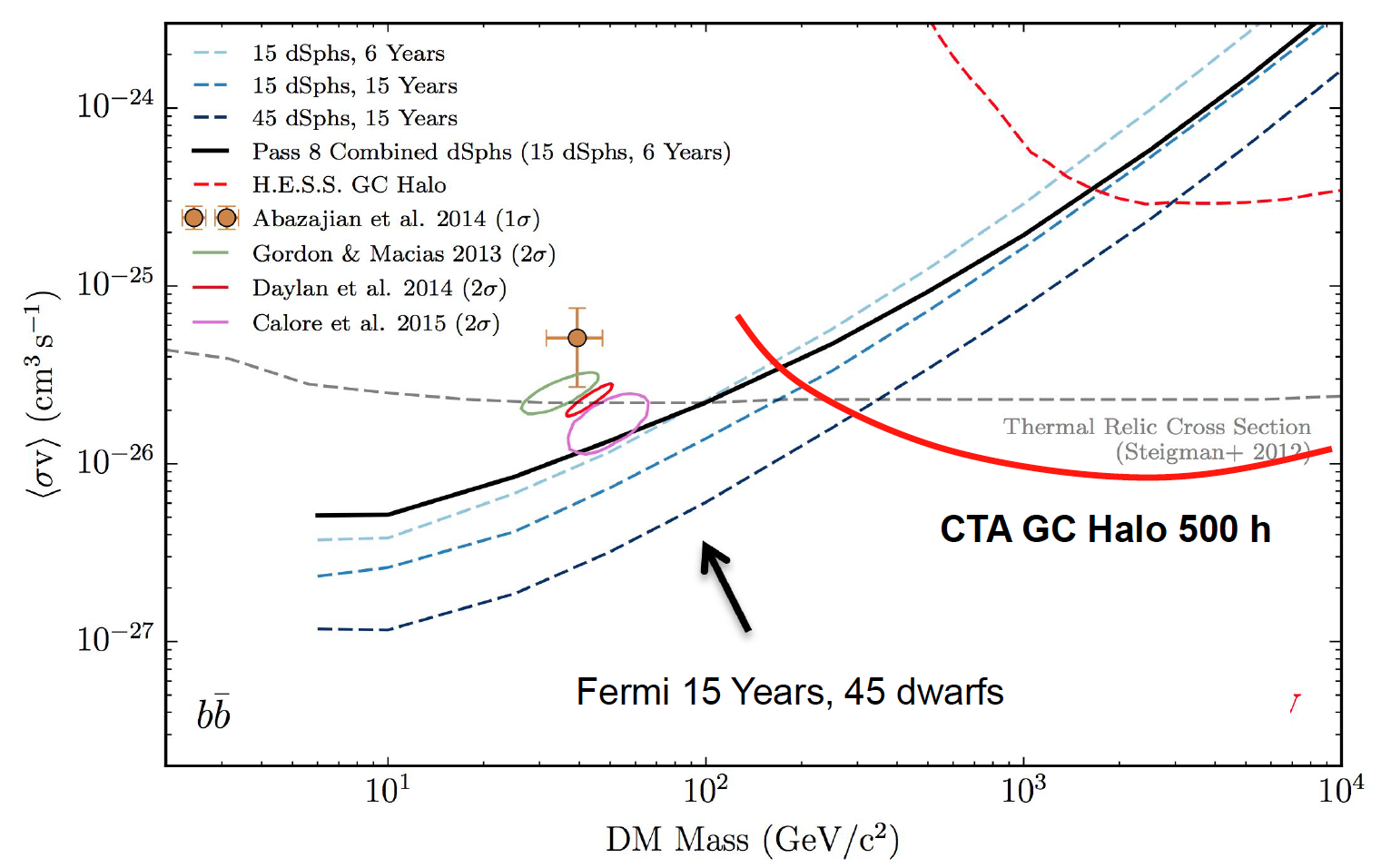}
        \caption{Expected CTA dark matter sensitivity by 2023. (Details see~\cite{ICRC921_CTA}).}
        \label{CTA}
\end{figure}

In the future the Cherenkov Telescope Array~(CTA) is expected to significantly enhance the dark matter sensitivity to masses between 100~GeV to 10~TeV. CTA is a next generation ground based gamma-ray observatory that will consist of more than 100~telescopes distributed over two site.  A southern site is located near Paranal (Chile), while the northern site is at La Palma on the Canary Islands (Spain). The Southern site will have 4~large-size telescopes (LST), 25~medium-size telescopes (MST), and 70~small-size telescopes (SST), while the Northern site will have 4~LST and 15~MST. Telescopes will have diameters of 23~m, 12~m, and 4~m, for LST, MST, and SST, respectively. 
CTA will have $\gamma$-ray sensitivity over the energy range of 20~GeV to 300~TeV, with a sub $0.1^\circ$ angular resolution and $7^\circ$ FOV. The expected dark matter sensitivity by 2023 is given in figure~\ref{CTA} and described in~\cite{Hutten:2016jko,ICRC921_CTA}.

Gamma-ray searches with space based telescopes is expected to continue and be expanded in scope as for example with the All-sky Medium Energy Gamma-ray Observatory (AMEGO). The AMEGO mission concept is designed to explore the MeV sky~\cite{AMEGO}. AMEGO will have 2.5~sr field of view and sensitivity for an energy range of 0.2~MeV to 10~GeV. It will combine three capabilities in MeV astrophysics: (1) continuum spectral studies, (2) polarization, and (3) nuclear line spectroscopy.

\subsection{Neutrinos}

Searches for dark matter have been carried out at neutrino telescopes, IceCube, ANTARES, Lake Baikal, and with the Super-Kaminokande neutrino detectors. 
The IceCube neutrino telescope consists of more than 5000~digital optical sensor modules (DOMs) distributed over 86~strings buried between 1.5~km to 2.5~km depth in the Antarctic ice at the geographic South Pole. The detector instruments a volume of about 1~gigaton. ANTARES consists of 12, 450~m long, detector lines at a depth of 2.5~km. Each line comprises 25 storeys with three 10-inch PMTs per storey. The PMTs are housed inside pressure-resistant glass spheres.
Neutrinos are detected through the Cherenkov light signatures of the relativistic charged particles created in neutrino interactions. Super-Kamiokande is a 50,000~ton  (22,500~ton fiducial) ring-imaging water Cherenkov detector located at a depth of 2.7~km water equivalent in the Kamioka Mozumi mine using 11K~20-inch~PMTs.

\subsubsection{Dark Matter Annihilation}

Latest results on the search for self-annihilating dark matter by IceCube~\cite{ICRC_Flis} and ANTARES~\cite{ANTARES_DarkMatter} are shown in figure~\ref{IACT_self_ann}. An effort to combine analyses from both experiments is also underway and has resulted in improved sensitivities~\cite{IceCubeANTARES} for a dark matter mass range of 100~GeV/c$^2$ to 500~GeV/c$^2$.

\subsubsection{Dark Matter Decay}

Interest in heavy dark matter decay scenarios has grown with the observation of PeV neutrino events by IceCube and several recent experimental searches have now been conducted. Heavy decaying dark matter has been suggested to explain the observed events in particular two neutrino events with PeV energy was intriguing. Such a line feature could be explained in models with singlet fermion in an extra dimension, hidden sector gauge boson, or gravitino dark matter with R-Parity violation. The necessary lifetime of the dark matter particle would be of order $10^{28}$~s (or exceeding the age of the Universe by 10 orders of magnitude)(see for example~\cite{Feldstein:2013kka}). For example a decay of a dark matter particle in a neutrino and Higgs boson ($\chi \rightarrow \nu h$) would result in a neutrino line and neutrino continuum, respectively. Bounds on such scenarios have been derived~\cite{Murase:2012xs,Esmaili:2012us,Aisati:2015vma,Rott:2014kfa}. Gamma-rays play an important role in constraining the dark matter interpretation of the IceCube neutrinos~\cite{Murase:2015gea}.

Boosted dark matter scenarios in which a heavy dark matter particle, $\phi$ decays to a lighter stable dark matter particle, $\chi$, which is then highly boosted have also proposed to explain the observed PeV events~\cite{Kopp:2015bfa}. The boosted dark matter particle could interact in the detector volume via nuclear recoil events resulting in hadronic showers. Interestingly these showers might be identified and distinguished from electromagnetic showers through the detection of delayed light from neutron capture~\cite{Li:2016kra, ICRC_Anna_Steuer}.

There are two contributions to the neutrinos flux from DM decay: (1) Dark Matter decaying in the Galactic halo and (2) Dark Matter decaying at cosmological distances. The Galactic contribution is results in a neutrino flux anisotropy and has a neutrino spectrum following that of dark matter decay. The extra-galactic contribution is isotropic and follows and the neutrino spectrum red-shifted. Near by galaxy clusters might show significant contributions and are interesting targets for future dark matter decay searches~\cite{Murase:2012xs,Esmaili:2012us}.

\begin{figure}[htb]
        \centering
        \includegraphics[width=0.55\textwidth]{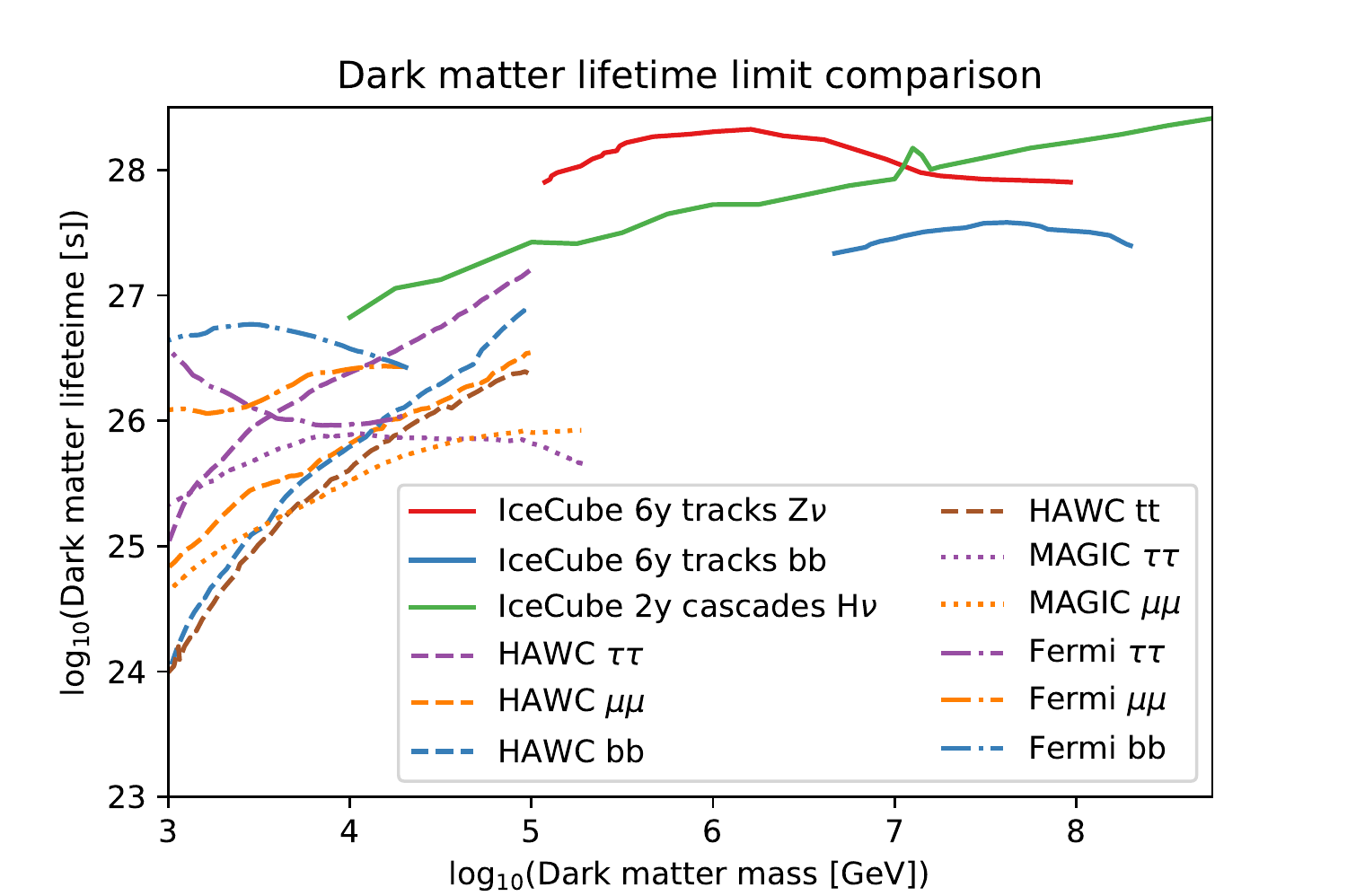}
        \includegraphics[width=0.40\textwidth]{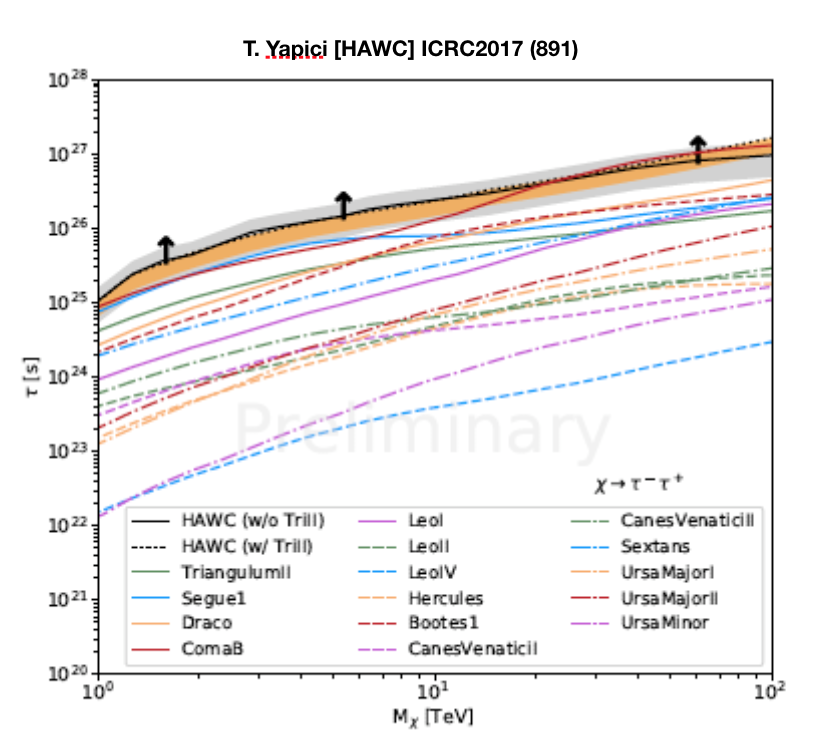}
        \put(-290,95){\large \textcolor{red}{Preliminary}}
        \caption{Left: Comparison of dark matter decay bounds from Fermi-LAT observations of the Galactic halo~\cite{Ackermann:2012rg}, MAGIC observation of Perseus~\cite{ICRC_MAGIC}, HAWC observations of dwarf spheriodal galaxies~\cite{ICRC_HAWC}, and IceCube observation of the Galactic halo and extra galactic diffuse fluxes~\cite{IceCube_DM_Decay}. Right: HAWC dark matter decay limit based on combined data of 15~dsph~\cite{ICRC_HAWC}.}
        \label{DM_decay}
\end{figure}

Latest results in the experimental searches for decaying dark matter are summarized in figure~\ref{DM_decay}. 95\% C.L. limits have been obtained by the MAGIC collaboration based on 270~h observations of the Perseus cluster taken during 2009 - 2017~\cite{ICRC_MAGIC}. 
Limits also exist from Veritas~\cite{Aliu:2012ga} based on Segue~1 observations and Fermi-LAT observations of the Milky Way halo~\cite{Ackermann:2012rg}. 
Two analyses have been conducted using data from the IceCube neutrino telescope. The first analysis used 6~years of muon-neutrino
data from the northern hemisphere, while the second analysis uses 2~years of cascade data from the full sky~\cite{IceCube_DM_Decay}.

\subsubsection{Solar Dark Matter}

Dark Matter particles from the Milky Way DM halo could be gravitationally captured by the Sun and accumulate in its centre. The capture process and standard calculations are given in~\cite{Gould:1987ir,Gould:1991hx,Wikstrom:2009kw,Danninger:2014xza}. Dark matter capture is initiated when a dark matter particle elastically scatters of a proton or nuclei to could loose enough energy to fall below the escape velocity of the Sun at the given radius. The dark matter particle hence becomes gravitationally bound to the Sun. In subsequent  collisions dark matter particles will settle to a thermal equilibrium in the Sun's core. In this way dark matter is expected to build up in the center of the Sun at a constant rate $\Gamma_{C}$, where it can then annihilate. The annihilation rate $\Gamma_{A}$ will increase with the amount of dark matter in the Sun until as much dark matter is captured as is lost due to annihilations. Once this equilibrium is reached the annihilation rate is independent of the dark matter self-annihilation cross section, but only depends on the scattering cross section, $\sigma$, of dark matter with the solar nuclei that initiates the capture in the Sun. As the Sun is a large hydrogen target the processes can be very sensitive to the spin-dependent dark matter scattering. 

Thermalization time scales for dark matter capture in the Sun has been evaluated with effective theories~\cite{ICRC916}. Using Monte-Carlo integration of WIMP trajectories and WIMP-nucleon interaction operators of a non-relativistic effective field theory the thermalization time scales and thermal profiles in the Sun were evaluated for isoscalar (isoscalar coupling WIMPs interact the same with protons and neutrons) and isovector (proton and neutron interactions have opposite signs) scenarios. With the exceptions of some fine-tuned cases, the thermalization time is significantly shorter than the age of the solar system and there is no significant thermal profile dependence except on the dark matter particle mass~\cite{ICRC916}. 
Dark matter capture in the Sun is also found to be very robust against any astrophysical uncertainty in the underlying dark matter velocity distribution~\cite{Danninger:2014xza,Choi:2013eda}.

\begin{figure}[htb]
        \centering
        \includegraphics[width=0.49\textwidth]{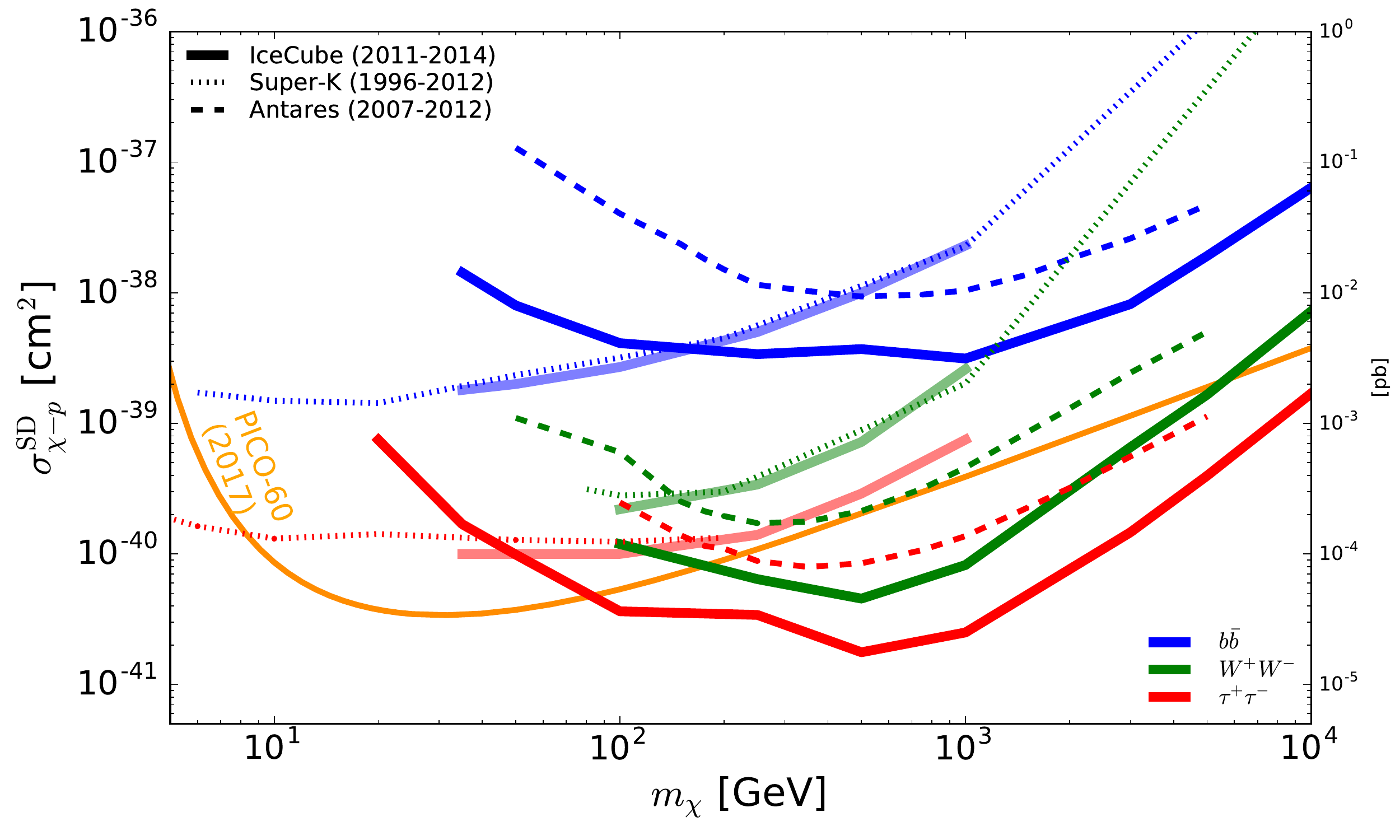}
        \includegraphics[width=0.49\textwidth]{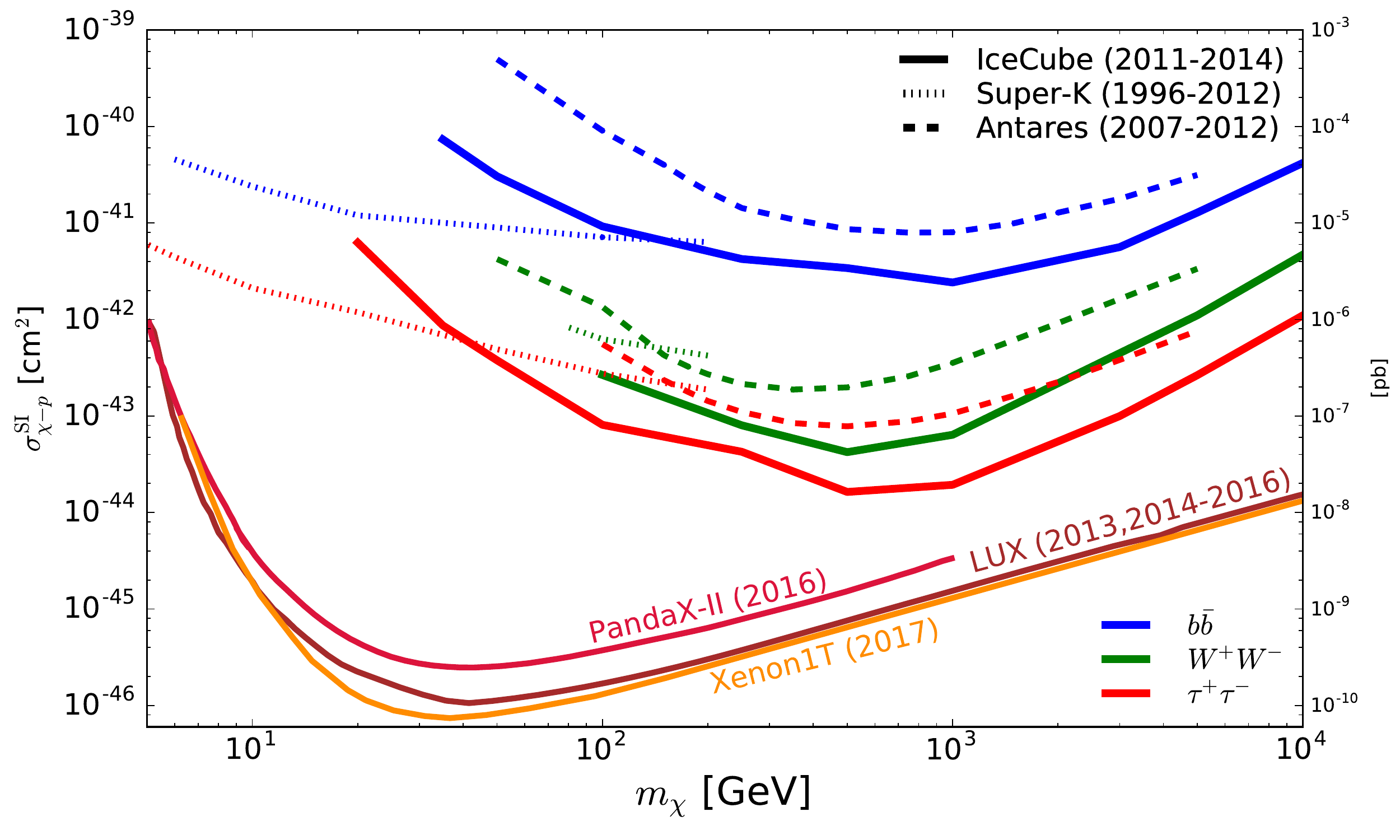}
        \put(-330,110){\textcolor{red}{Preliminary}}
        \put(-150,110){\textcolor{red}{Preliminary}}
        \caption{Current limits on the spin-dependent (left) and spin-independent (right) scattering cross sections of dark matter with protons. Results from indirect searches are compared with direct searches. Faint lines indicate the results of an all neutrino flavor based IceCube analysis (Details see~\cite{ANTARES_DarkMatter,SolarWIMP_Jin}).}
        \label{SolarWIMP}
\end{figure}

Searches for solar dark matter have been performed by various neutrino detectors and neutrino telescopes~\cite{Danninger:2014xza}. The most stringent bound on dark matter annihilation in the Sun have been performed by ANTARES~\cite{ANTARES_DarkMatter,Adrian-Martinez:2016gti}, IceCube~\cite{Aartsen:2016zhm,SolarWIMP_Jin}, and Super-Kamiokande~\cite{Choi:2015ara}. Current bounds on the spin-dependent ($\sigma^{\rm SD}$) and spin independent scattering cross sections ($\sigma^{\rm SI}$) of dark matter with nucleons are given in figure~\ref{SolarWIMP}. Limits from neutrino telescopes depend on the neutrino spectrum produced in the self-annihilation. As benchmark channel dark matter annihilation into $b\bar{b}$ and $\tau^{+}\tau^{-}$ are shown. Limits are compared to those from direct detection experiments. Strong bounds from the PICO-60 C$_3$F$_8$ bubble chamber dark matter direct detection experiment in Snolab exclude a 30~GeV WIMP with spin-dependent WIMP-proton cross sections above 3.4$\times 10^{-41}$~cm$^2$ using 1167-kgday exposure~\cite{Amole:2017dex}.

\subsubsection{Solar Atmospheric Neutrinos}

Cosmic rays interacting in the solar atmosphere can produce high-energy neutrinos that form a natural background to dark matter searches from the Sun. This solar atmospheric neutrino flux is an irreducible background for Solar DM searches and forms a sensitivity floor that is reached when the expected event rates from solar atmospheric neutrinos equals that of DM-induced neutrinos~\cite{Arguelles:2017eao,Ng:2017aur,Edsjo:2017kjk,Masip:2017gvw}. While the DM induced neutrino spectrum strongly depends on the dark matter mass, ${\rm m}_{\chi}$, and annihilation channels, the energy spectrum of solar atmospheric neutrinos is more predictable. Dark matter searches below the solar atmospheric neutrino floor are possible, however their sensitivity strongly depends on the ${\rm m}_{\chi}$ and annihilation channels. In general it should be noted that solar WIMP spectra are strongly attenuated above a few hundred GeV due to absorption in the Sun, while solar atmospheric neutrino spectrum is expected to extend to TeV energies. The high energy neutrino flux is also what makes the solar atmospheric neutrino flux most detectable in searches for IceCube~\cite{Sun_Jin} and ANTARES~\cite{ANTARES_Sun}. IceCube sensitivity are encouraging and a detection in the future might be possible~\cite{Sun_Jin}.


\subsubsection{Dark matter captured in the Earth}

\begin{figure}[htb]
       \centering
        \includegraphics[width=0.43\textwidth]{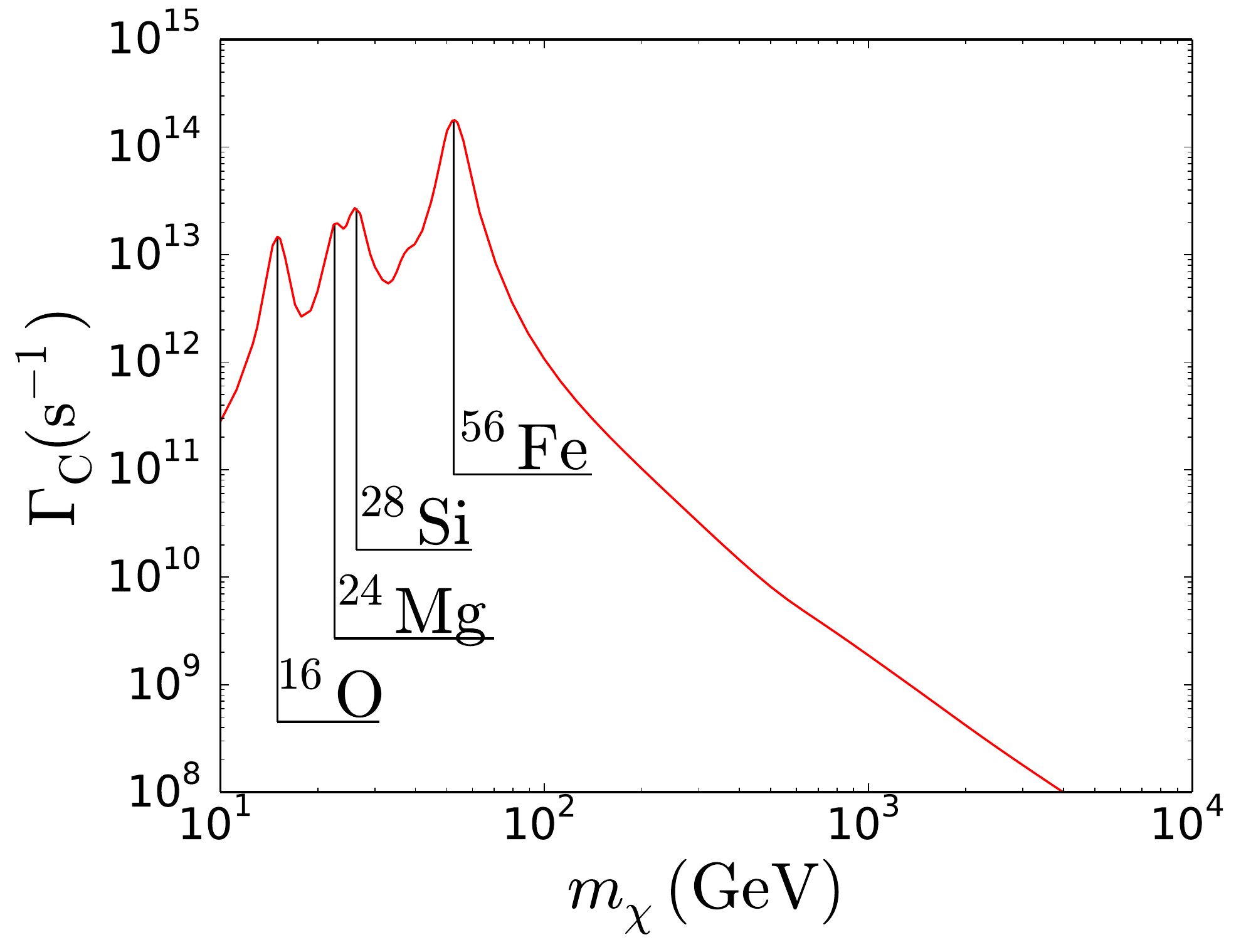}
        \includegraphics[width=0.46\textwidth]{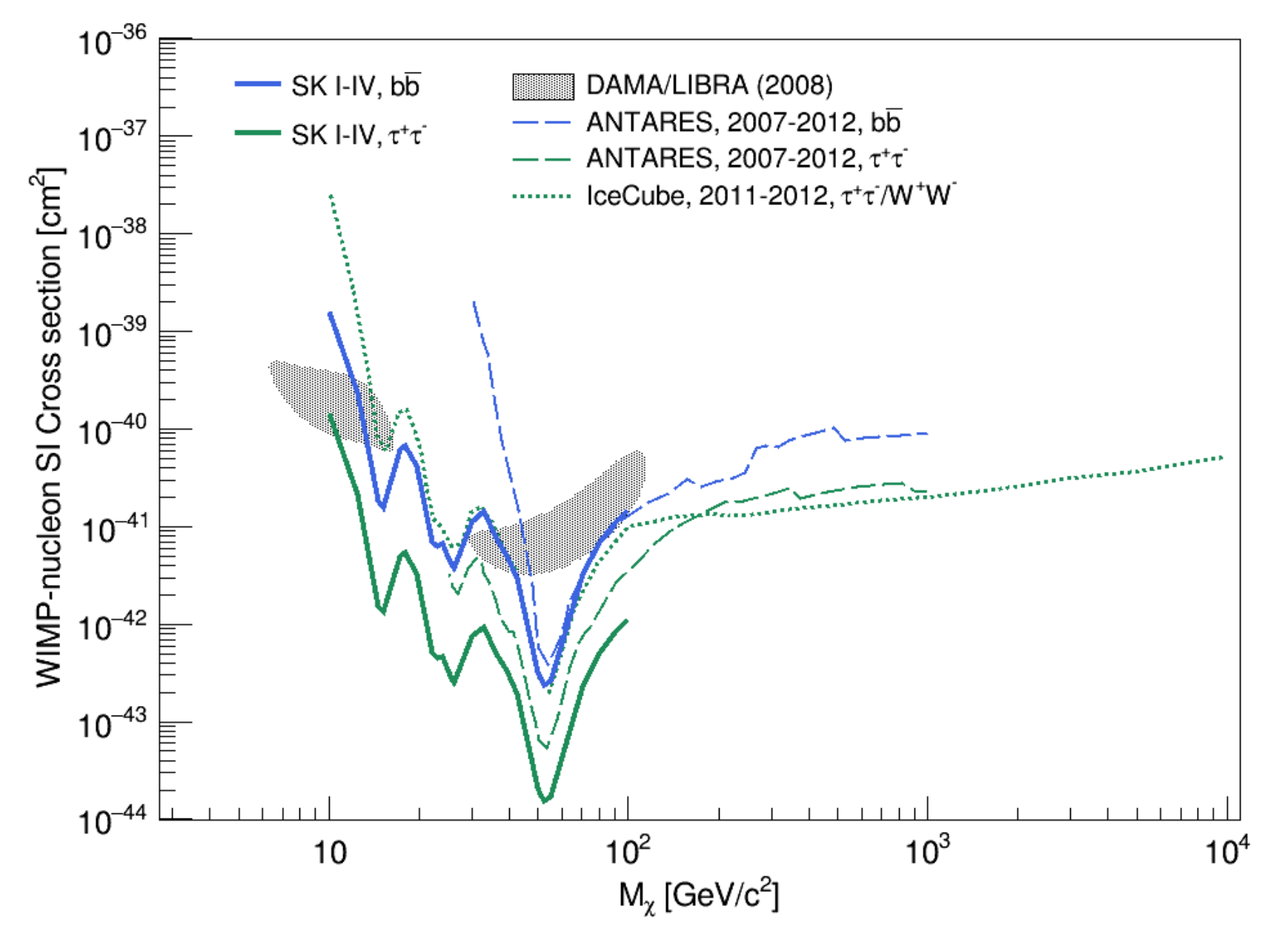}
        \caption{Left: Expected dark matter capture rates in the Earth for $\sigma^{\rm SI} = 10^{-44} {\rm cm}^2$~\cite{ICRC_IceCube_Earth}. Right: Indirect bounds on dark matter annihilations in the center of the Earth from Super-K (preliminary results~\cite{WIN2017_SK}), ANTARES~\cite{ANTARES_DarkMatter,Albert:2016dsy}, and IceCube~\cite{ICRC_IceCube_Earth,Aartsen:2016fep}.}
        \label{EarthWIMP}
\end{figure}

Dark matter could be captured in the Earth and annihilate resulting in a potentially detectable excess in vertical up-going neutrino events. Contrarily to the Sun, the Earth is expected not to be in equilibrium between capture and annihilation making any signal prediction highly uncertain. Capture is strongly enhanced if $m_{\chi}$ closely matches the nuclear mass of an abundant element in the Earth. The search for DM captured in the Earth is experimentally more challenging compared to Solar DM searches as no off-source region can be defined. A summary of the most recent experimental searches is given in figure~\ref{EarthWIMP}.

\subsubsection{Prospects with next generation neutrino detectors and telescopes}

Sensitivity to $\sigma^{\rm SD}$ for dark matter masses below 100~GeV can be significantly improved with neutrino telescopes ORCA and PINGU in solar dark matter searches~\cite{ANTARES_DarkMatter,SolarWIMP_Jin}. Hyper-Kamiokande has sensitivity in the same mass range with the additional capability to test scenarios in which dark matter predominantely annihilates to light quarks~\cite{Rott:2012qb,Bernal:2012qh,Rott:2015nma}. KM3NeT or IceCube-Gen2 are expected to effectively test high mass dark matter decay and other beyond the SM scenarios~\cite{Rasmussen:2017ert}.


\subsection{Anti-matter and electron/positrons}

Searches based on electron/positron signals as well as anti-matter are discussed. The positron excess is discussed separately in the Anomalies section.

\begin{figure}[htb]
        \centering
         \includegraphics[width=0.483\textwidth]{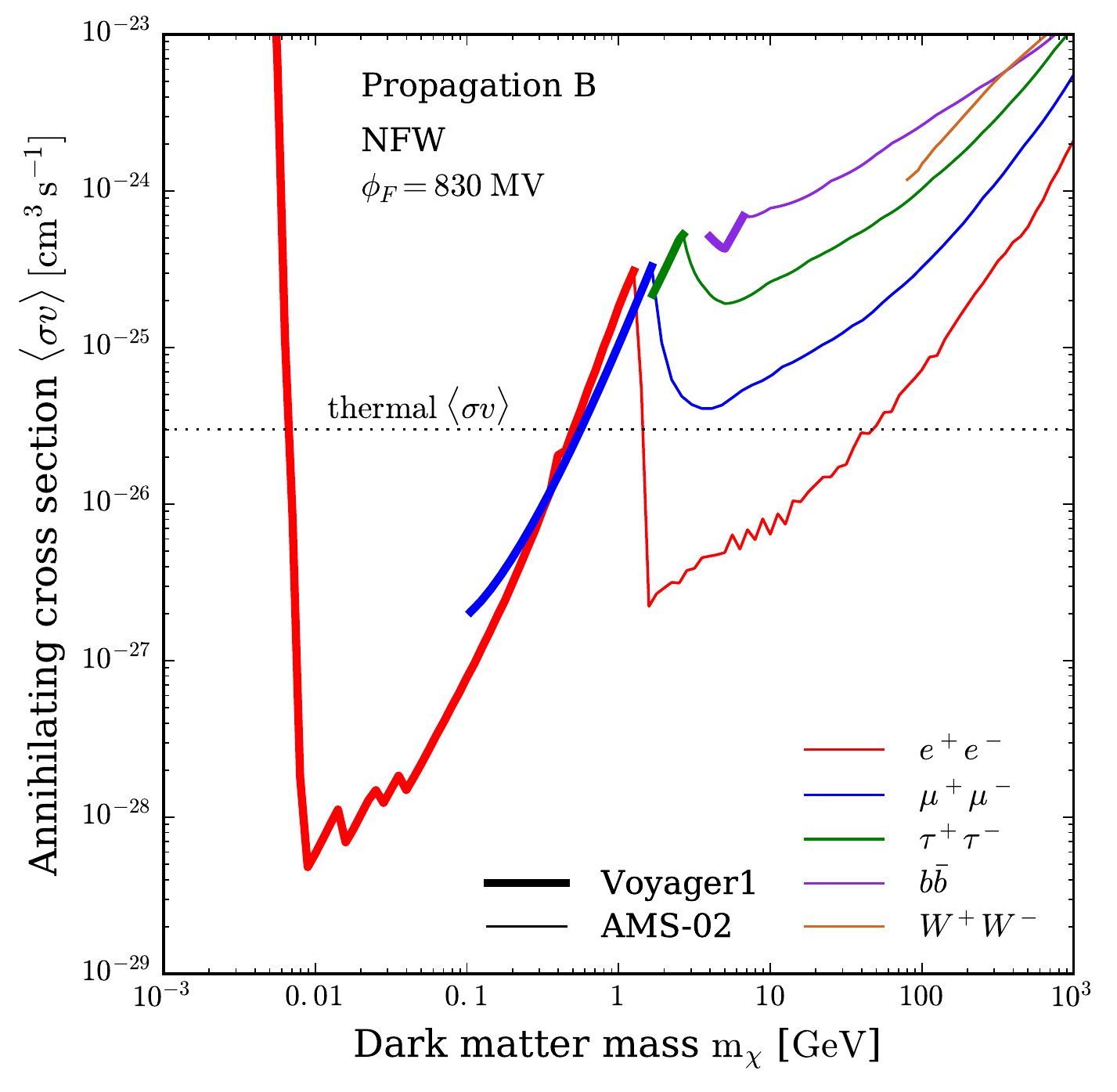}     
        \includegraphics[width=0.49\textwidth]{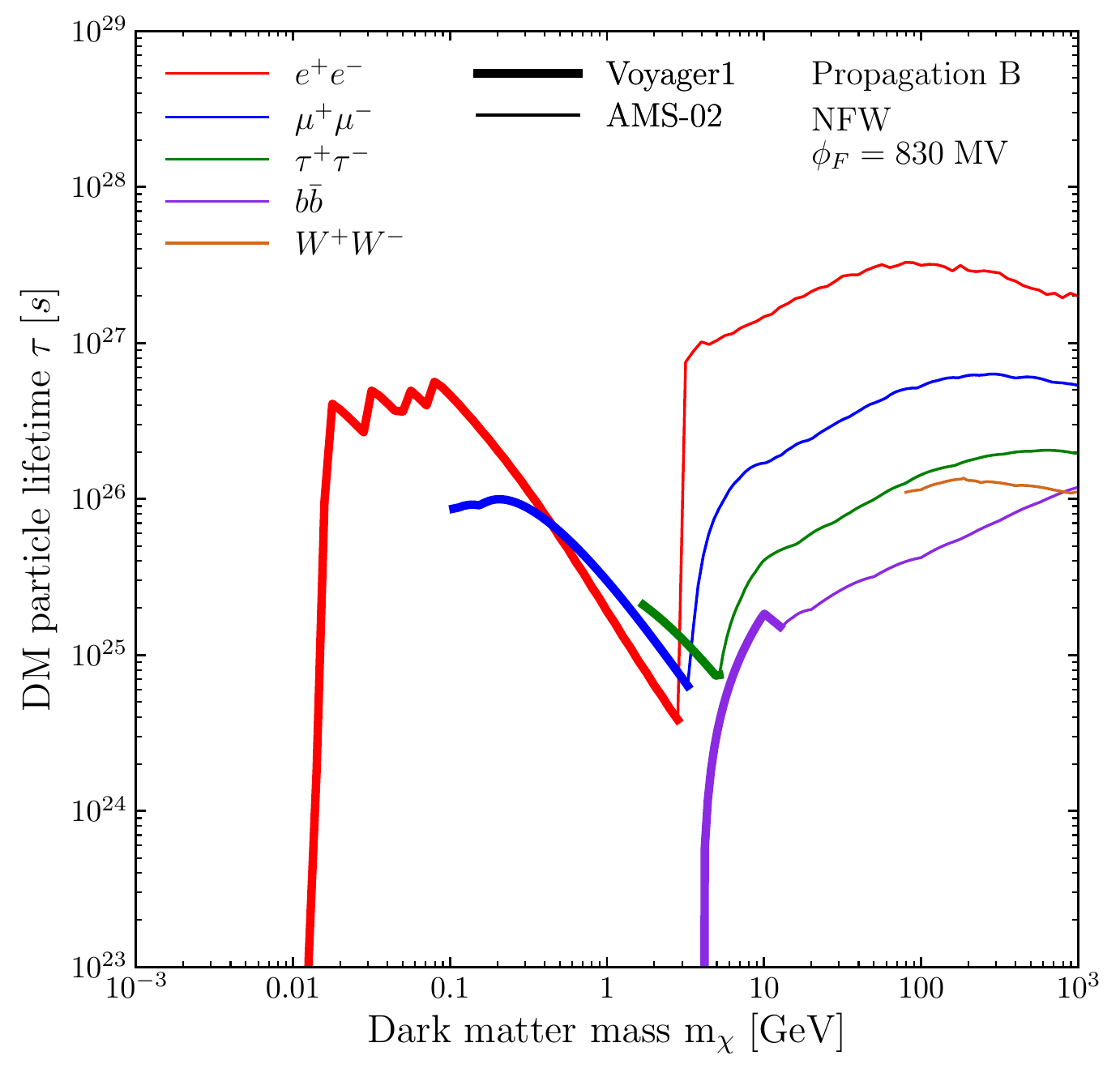}
        \caption{Bounds on dark matter annihilation and decay derived from Voyager-I and AMS-02 data~\cite{voyager_ref}. (Figures from~\cite{voyager_ref})}
        \label{voyager}
\end{figure}

Indirect searches have largely focused on dark matter with GeV - TeV scale dark matter. Below a GeV only a few channels are kinematically allowed. These include $\pi$, $\mu^{+} \mu^{-}$, e$^{+}$ e$^{-}$, photons, and neutrinos.
Interstellar sub-GeV electrons and positrons are shielded by the heliosphere and hence put them out of reach of most detectors. With the exception of Voyager-I spacecraft that has crossed the heliopause during summer 2012. Constraints on DM can be derived requiring that observed electron and positron fluxes below and above a GeV do not exceed the Voyager-I and AMS-02 data, respectively~\cite{voyager_ref}. Bounds are shown in figure~\ref{voyager}.

\begin{figure}[htb]
        \centering
         \includegraphics[width=0.95\textwidth]{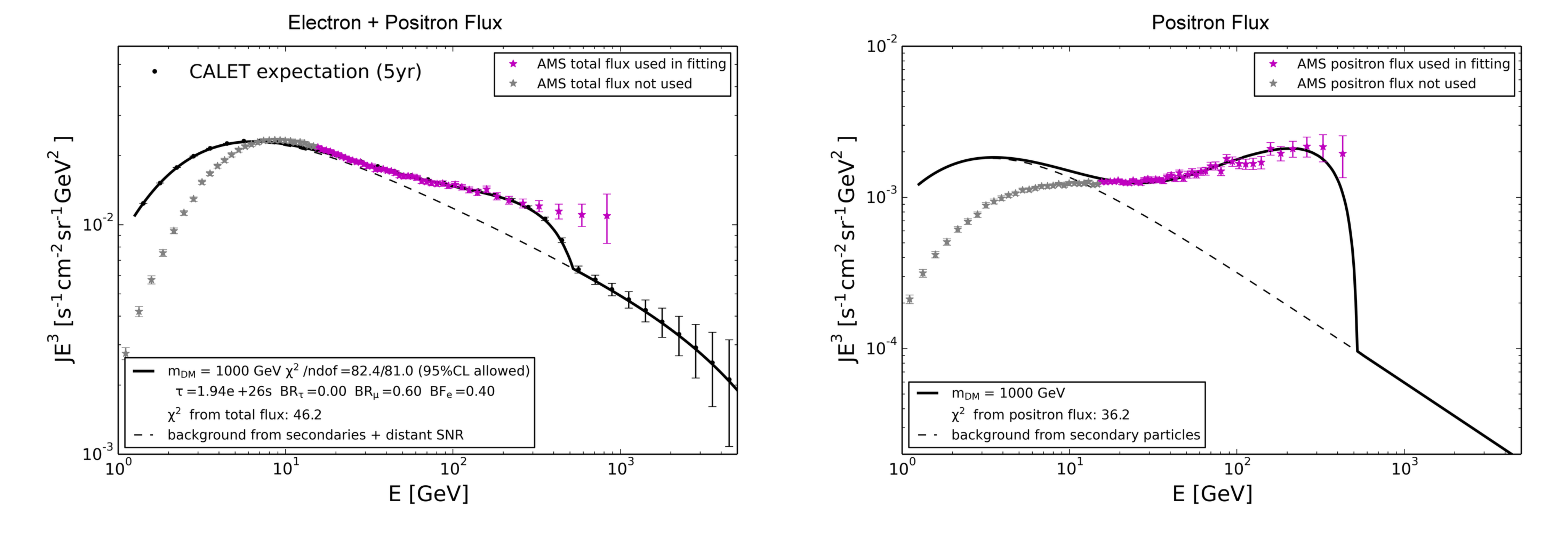}   
        \caption{Expected $e^{+} + e^{-}$ (left) and $e^{+}$ (right) spectra of a 1~TeV DM decay with branching fractions 0.6 $\mu\mu\nu$ and 0.4~$ee\nu$ on top of the background (dotted line) fitted to the measured by AMS-02 fluxes. The plot demonstrates how 5~yrs of CALET data is expected to improve the spectral measurement~\cite{CALET_919} (Figures from~\cite{CALET_919}).}
        \label{fig_calet}
\end{figure}

The ISS-based CALorimetric Electron Telescope (CALET), which is in operation since October 2015 can study the ($e^{+} + e^{-}$) spectrum in the TeV region with 2\% energy resolution and little background contamination (1:$10^5$ proton rejection). CALET is of particular interest to detect a sudden drop in the electron spectrum that could help distinguish pulsar models from dark matter induced fluxes~\cite{Motz:2015cua}. CALET can also test fermionic dark matter models that could fit the observed AMS-02 lepton spectrum and at the same time evade strict constraints from gamma-ray observations. Dark matter decay to 2 charged and 1 neutral leptons ($ee\nu$, $\mu\mu\nu$, $\tau\tau\nu$) remain of interest and are a prime target for CALET. Fig~\ref{fig_calet} demonstrates the expected uncertainty on the electron/positron flux that can be achieved with 5~years of CALET data~\cite{CALET_919}. DAMPE is working on an electron spectrum up to 5~TeV and H.E.S.S. has already presented preliminary result on the total electron-positron flux above 1~TeV~\cite{HESS_ep_ref}.

The GAPS~\cite{ICRC_GAPS} experiment is designed search for dark matter by measuring
low-energy cosmic ray antideuterons and antiprotons. GAPs detects antideuterons and antiprotons using an exotic atom technique, which distinguishes itself from experiments like BESS and AMS that use magnetic spectrometers. After a successful test flight of a prototype instrument from Taiki, Japan in 2012, GAPS has been approved by NASA to proceed towards the full science instrument. The first  long-duration balloon in Antartica is expected in late 2020.


\section{Anomalies}

\subsection{Sterile Neutrino -- 3.5~keV Line}

A recent 3.5~keV x-ray line observed from various targets may indicate the existence of 7~keV sterile neutrino~\cite{Abazajian:2017tcc}. 
A $4-5~\sigma$ significance detection of the 3.5~keV x-ray line in stacked observations of 73 galaxy clusters with the MOS and PN spectrometers aboard XMM-Newton, as well as a consistent signal from the Perseus cluster of galaxies observed with the Chandra telescope was reported in 2014~\cite{Bulbul:2014sua}. A consistent signal was found from the Andromeda galaxy and the Perseus cluster using data from the XMM-Newton satellite~\cite{Boyarsky:2014jta}. 
The 3.5~keV signal should however be regarded with caution as the signal is not consistently observed. Several observations rule out the signal, these include observations of Draco with XMM~\cite{Jeltema:2015mee}. Hitomi (formerly known as Astro-H) was ideally suited to study the 3.5~keV line, however the satellite could not acquire enough data during its very short life time in orbit. With approximately 7\% of the observation time needed to confirm the excess, Hitomi observed a dip from the Perseus Galaxy cluster and a limit was placed~\cite{Aharonian:2016gzq}.

\begin{figure}[htb]
        \centering
        \includegraphics[width=0.53\textwidth]{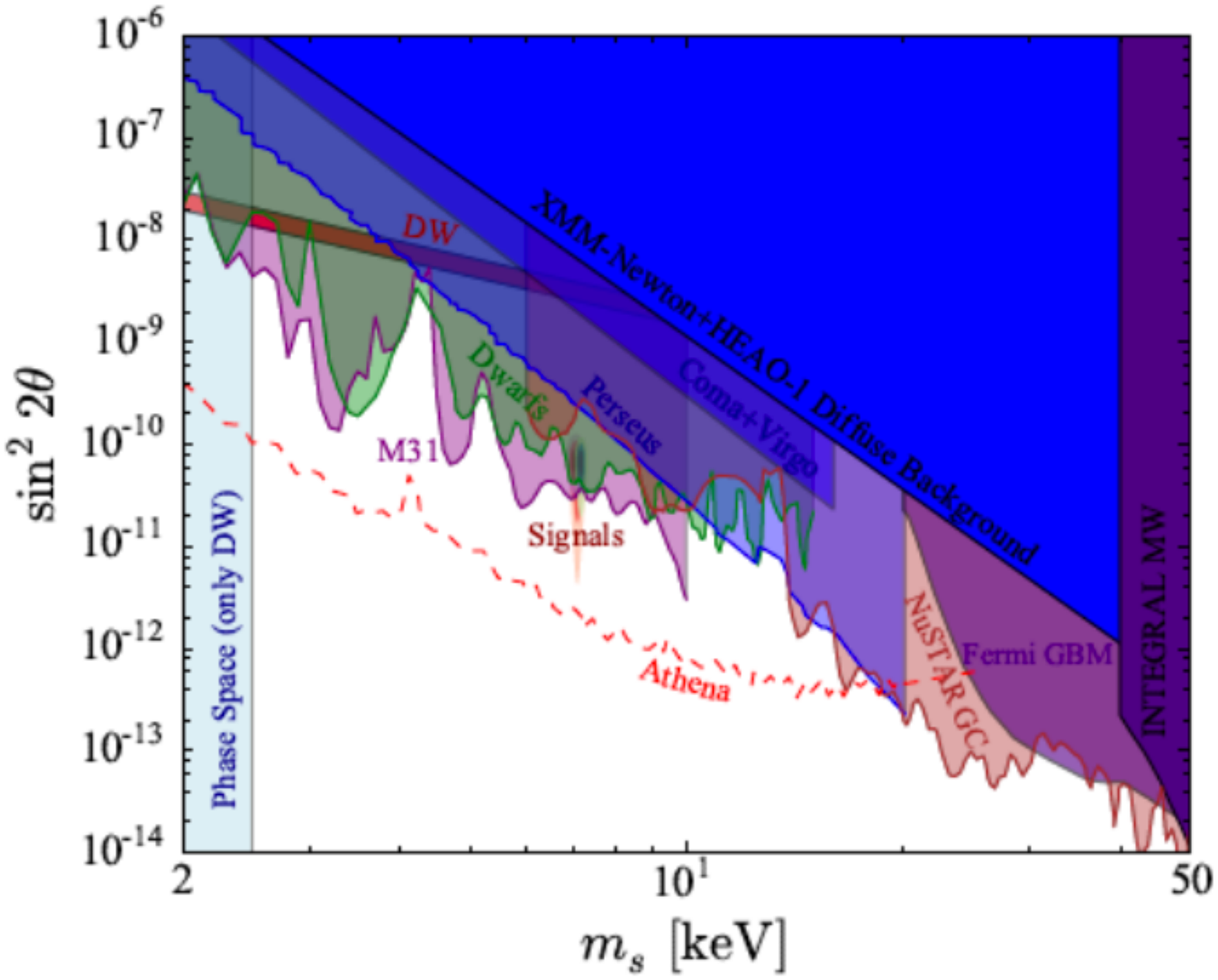}
        \includegraphics[width=0.45\textwidth]{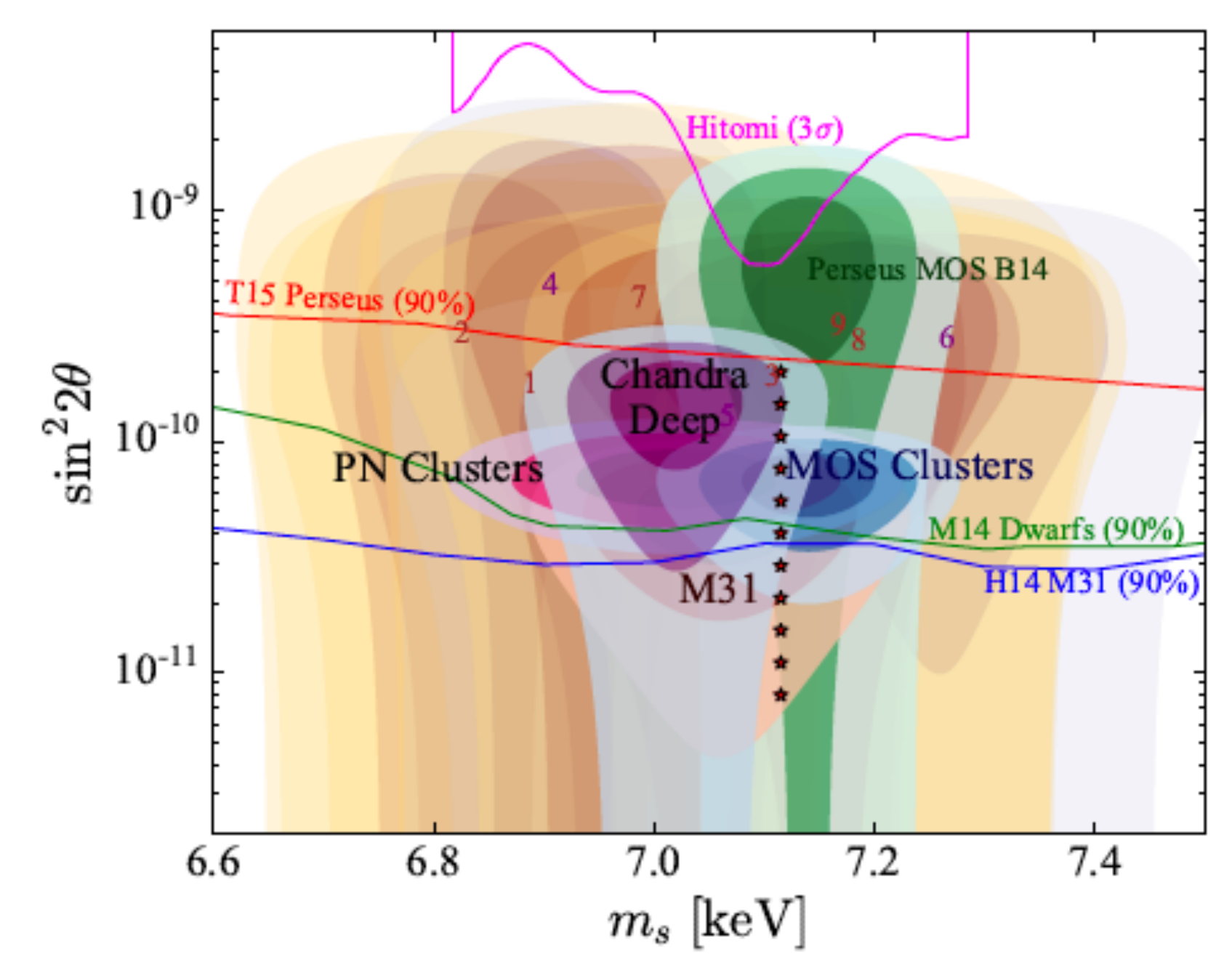}
        \caption{Left: Bounds on sterile neutrinos as function of the mass $m_{s}$. Right: Signals of 3.5~keV line. Details see~\cite{Abazajian:2017tcc} (Figures from~\cite{Abazajian:2017tcc}). }
        \label{3.5kev_sterileneutrino}
\end{figure}

The 3.5~keV line could also be due to astrophysical backgrounds.
The line could also be due to atomic transitions of highly-ionized atoms or through charge exchange (CX)~\cite{Gu:2015gqm}.
Highly-ionized potassium (K~XVIII) has two lines in the close vicinity of 3.5~keV, that are expected to be indistinguishable from a 3.5~keV in current data. CX between bare sulfur and neutral hydrogen interacting with a relative velocity of 200 km/s can produce a line at 3.44~keV (This is the S~XVI CX line). 

The final word on the origin of the 3.5~keV line is still out. Good spectral resolution is the key to confirm the line and to distinguish a sterile neutrino signal from astrophysical backgrounds. In the future data from ATHENA is expected to provide sufficient sensitivity and energy resolution to study the 3.5~keV line. NASA and JAXA have also announced plans to refly Hitomi in 2021. Dark matter velocity spectroscopy is a promising technique that could be used to distinguish a sterile neutrino signal from astrophysical background but requires an energy resolution of $0.1\%$~\cite{Speckhard:2015eva}. Micro-X data might be suitable to test this new method~\cite{Powell:2016zbo}.

\subsection{Galactic Center}

An excess of GeV gamma-rays in Fermi-LAT data was first reported by Goodenough and Hooper~\cite{Goodenough:2009gk}. The Fermi-LAT collaboration reported that the spectrum and morphology of the excess is sensitive to the assumed diffuse emission model. However the excess is still statistically significant under all models tested~\cite{TheFermi-LAT:2015kwa,TheFermi-LAT:2017vmf}. The excess has bee interpreted as originating from self-annihilating dark matter. Under this assumption depending on the annihilation channel typically dark matter masses are approximately 10~GeV ($\tau^{+}\tau^{-}$) or 35~GeV ($b\bar{b}$)~\cite{Calore:2014nla}, with annihilation cross sections close to that of the thermal relic cross section~\cite{Steigman:2012nb}. The dark matter scenarios are inconsistent or at best in tension with limits from dwarf spheriodal galaxies.
Recently, mounting evidence for large contribution from pulsars~\cite{PabloM.SazParkinsonfortheFermiLAT:2017vip,Lee:2014mza, Fermi-LAT:2017yoi,Bartels:2015aea}, could offer an explanation for the excess. A more recent analysis of the 2FIG catalog does not provide significant support for either a pulsar or a dark
matter interpretation of this signal~\cite{Bartels:2017xba}.

A new template fit based on energy spectra for each possible process of gamma-ray emission finds that gamma-ray production from $\pi^0$ in molecular clouds could explain the Galactic Center excess~\cite{ICRC908,deBoer:2017sxb}. The study relied on a CO Skymap, as tracer for molecular clouds, from the Planck Satellite through the measurement of CO rotation lines~\cite{ThePlanck:2013dge}. The Galactic Center excess resembles in morphology more that of the central molecular cloud zone (CMZ)  instead of a spherical DM profile. Note that the Fermi-LAT Collaboration showed that the GC excess is robust to the inclusion of a CMZ template~\cite{TheFermi-LAT:2017vmf}.

\subsection{Positron Excess}

A large excess of positrons above 10~GeV inconsistent with secondary production expectations has been observed PAMELA and was later confirmed by Fermi-LAT. AMS-02 precisely measured the flux and confirmed the increase in ratio between $\phi(e^{+})/\phi(e^{+} + e^{-})$~\cite{ICRC_AMS_Kounine}. The excess could originate from self-annihilating or decaying dark matter and has been extensively discussed in the literature. The dark matter interpretation is however in tension with indirect bounds based on gamma-rays, neutrinos, and radio.

\begin{figure}[htb]
        \centering
        \includegraphics[width=0.34\textheight]{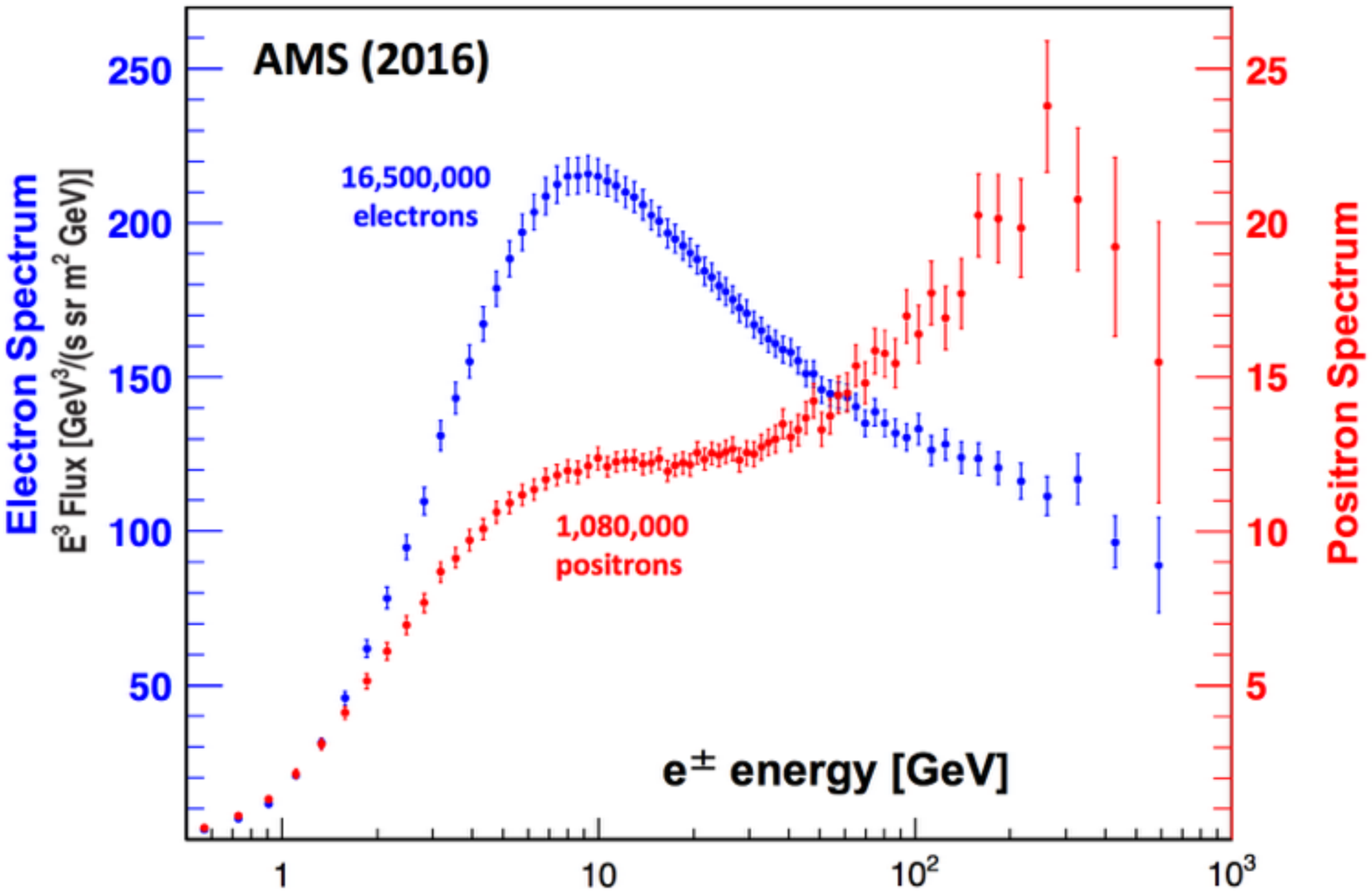}
        \includegraphics[width=0.46\textwidth]{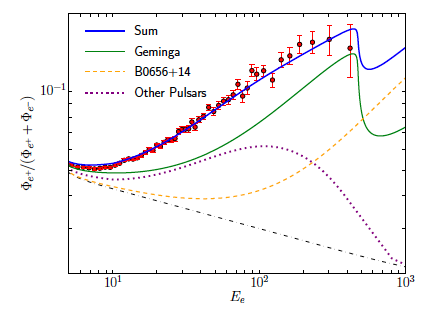}
        \caption{Left: Electron and positron flux~\cite{ICRC_AMS_Kounine} as observed by AMS-02. Right: Recent fit of the electron and positron flux to pulsar spectrum~\cite{Hooper:2017gtd}. (Figures from~\cite{Hooper:2017gtd})}
        \label{AMS_ep}
\end{figure}

Astrophysical sources such as pulsars potentially provide large signal contributions that could explain the observed spectra.  Measurements of the Geminga and B0656+14~(Monogem) pulsars by HAWC and Milagro indicate that these objects generate significant fluxes of very high-energy electrons~\cite{Hooper:2017gtd}.
Figure~\ref{AMS_ep} shows the observed electron and positron flux and a fit to pulsar models.
The increase in positron to electron ratio with energy remains of high interest, however any dark matter interpretation based on the electron and positron fluxes alone remains unconvincing.

\section{Direct Detection}

In dark matter direct detection currently the following trends are observed:
\begin{itemize}
\item Dark matter searches using liquid Nobel gases (such as XENON, LUX, PANDA-X, ...) are making rapid progress and are now moving towards multi ton scales. The next generation of these experiments (LZ, ...) and liquid argon based detectors (DARWIN, DarkSide, ... ) are expected to approach the neutrino floor.
\item Bubble chambers prove to be extremely competitive for spin-dependent searches (such as PICO, ...)
\item Cryogenic detectors (such as CRESST-III, CDMSLite, ...) have proven to be very effective in exploring the light DM region (m$_{\chi}<10$~GeV/c$^2$).
\item The long standing DAMA anomaly is expected to be confirmed or resolved in the near future with multiple new experiments coming on-line now that rely on the identical detector technology.  
\item Next generation dark matter experiments are becoming multi-purpose experiments. 
\end{itemize}

\begin{figure}[htb]
        \centering
        \includegraphics[width=0.48\textwidth]{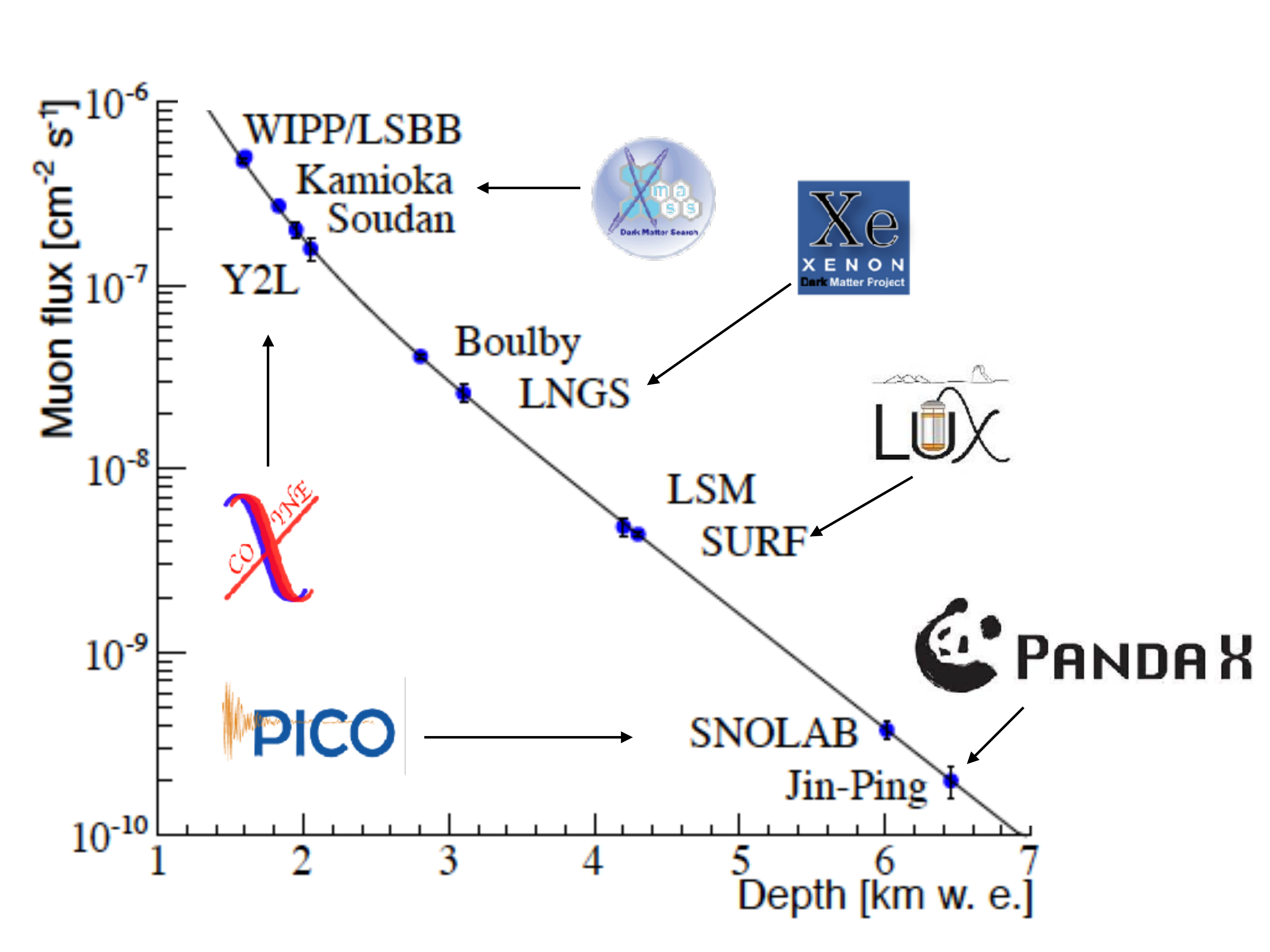}
        \includegraphics[width=0.48\textwidth]{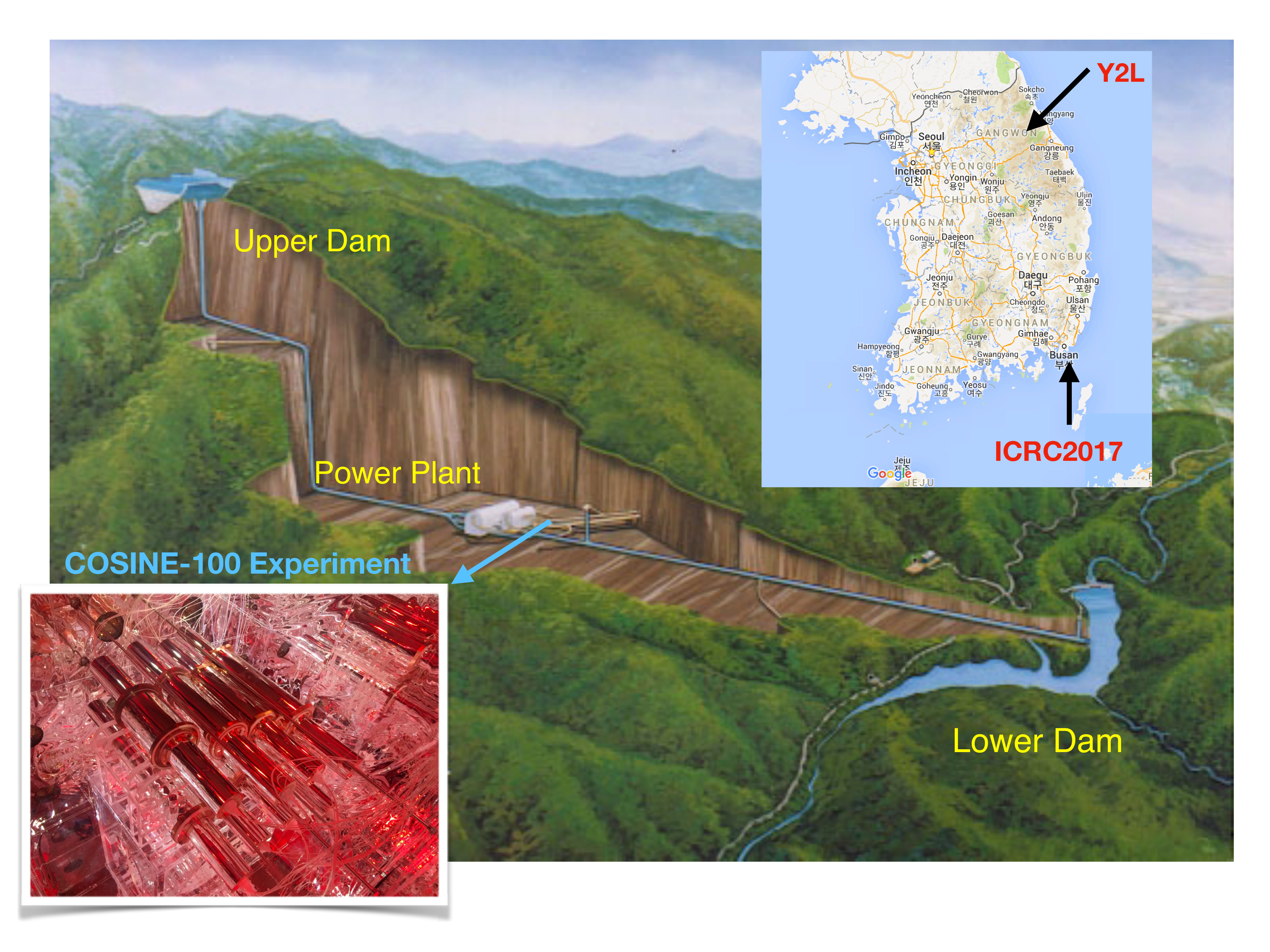}
        \caption{Left: Underground labs and selected experiments based on relevance to ICRC~2017 presentations.Right: Location of the COSINE-100 Experiment at Y2L.}
        \label{underground_labs}
        \label{Y2L}
\end{figure}

At ICRC~2017 the focus was on liquid Xenon based experiments and the COSINE-100 experiment, which is located in Korea. Figure~\ref{underground_labs} shows underground laboratories and selected experiments with relevance to conference presentations and their rock overburden in meter water equivalent (m.w.e.).

\begin{figure}[htb]
        \centering
        \includegraphics[width=0.32\textwidth]{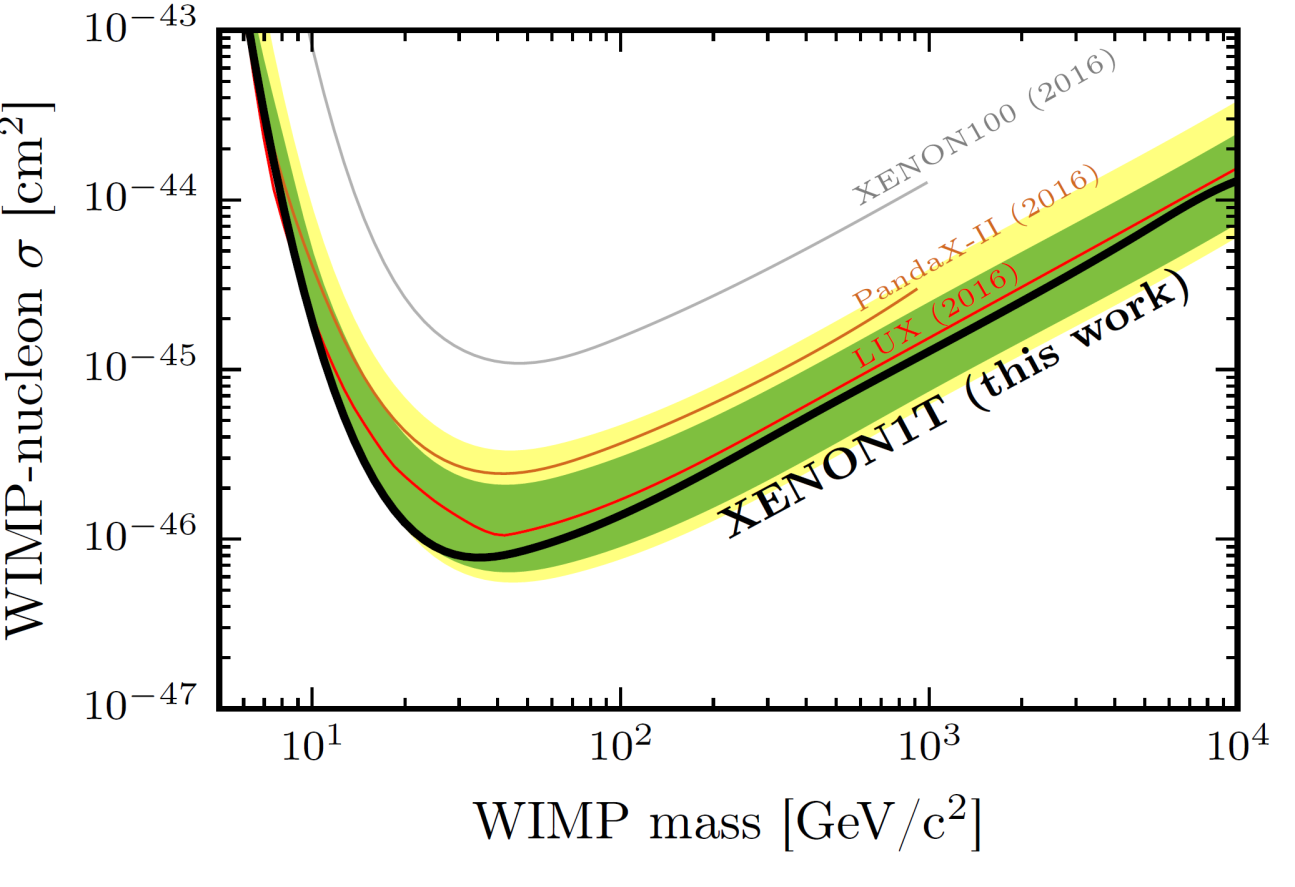}
        \includegraphics[width=0.32\textwidth]{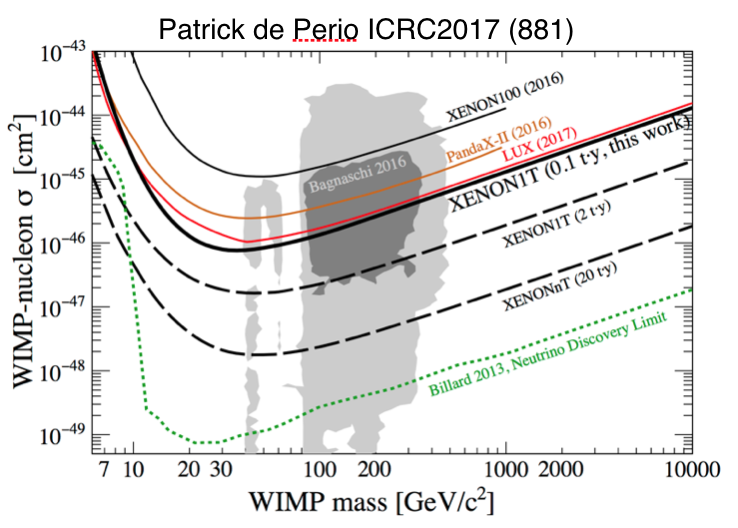}
        \includegraphics[width=0.32\textwidth]{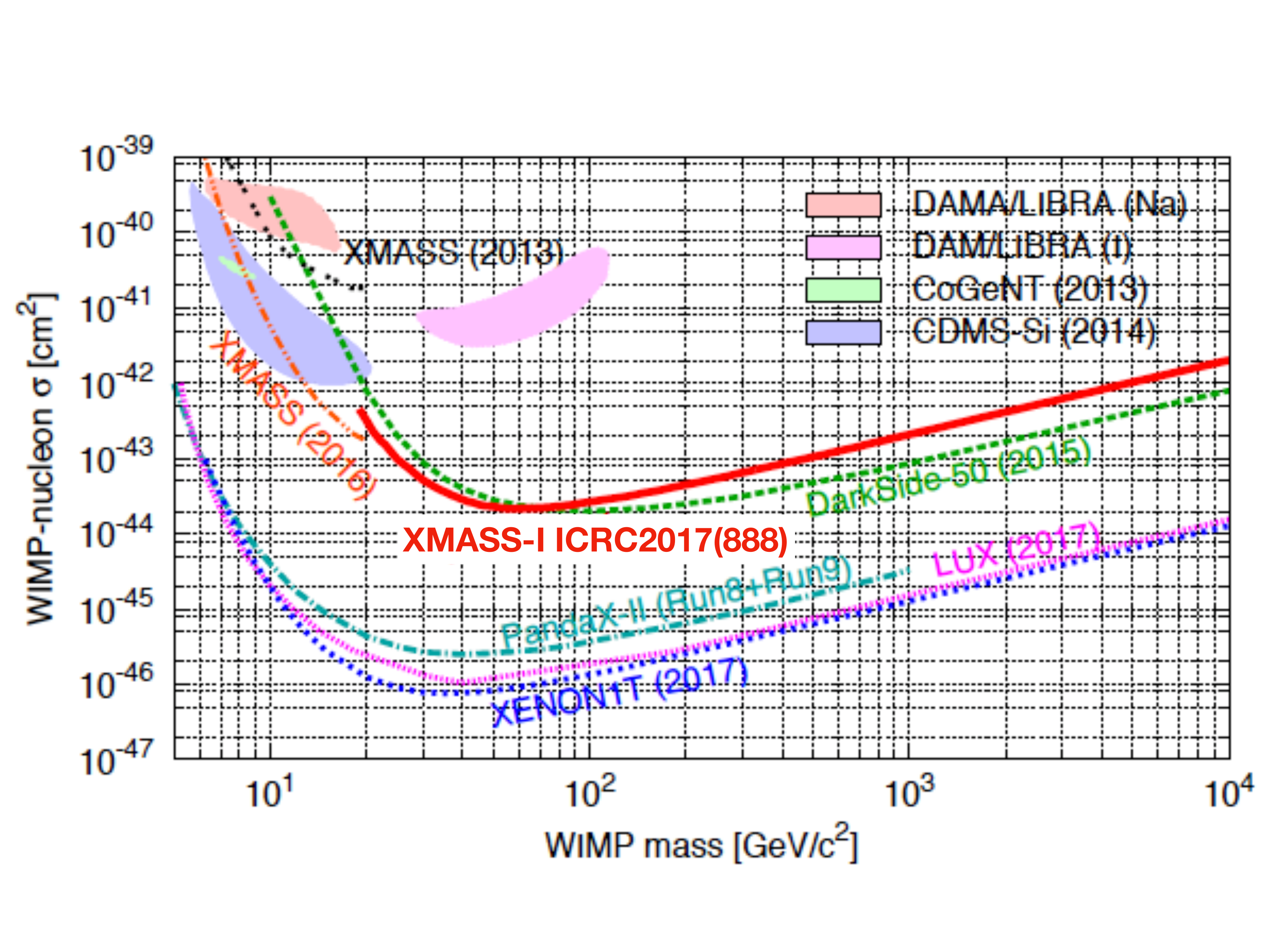}
        \caption{Left: Latest results from dark matter direct detection on the spin-independent elastic dark matter nucleon scattering cross section as function of the dark matter mass~\cite{XENON1T_32days,ICRC_XENON1T}. Middle: Expected sensitivity of the XENON experiment and proposed upgrades~\cite{ICRC_XENON1T} (figures from~\cite{XENON1T_32days,ICRC_XENON1T}). Right: Preliminary limits obtained from the XMASS experiment are shown. Details are given in~\cite{ICRC_XMASS_888} (figure from~\cite{ICRC_XMASS_888}).
        }
        \label{SI_limits}
\end{figure}

Liquid xenon (LXe) has proven to be an ideally suited target material, it offers high radiopurity, high density, which is beneficiary for shielding and provides a compact target volume. Dual phase xenon time projection chambers (TPC) detect scintillation light (S1) from nuclear recoil events and the ionization yield via a proportional scintillation signal (S2) from electrons accelerated in the electric field entering the gas phase. Signals are detected via PMTs mounted at the top and bottom of the detector. Electrons have a uniform drift velocity in liquid xenon, which allows the interaction depth of the event to be determined by measuring the time difference between the S1 and S2 signal. The ionization/scintillation ratio (S2/S1) allows for the rejection of electron recoil from nuclear recoil events.

The dual phase LXe technology has been successfully utilized for the XENON, LUX, PANDA-X experiments and resulted in limits on the spin-independent WIMP nucleon scattering cross section $\sigma^{SI}$ below $10^{-46} {\rm cm}^2$ for dark matter masses of 30~GeV/c$^2$. 

The XENON1T experiment is located at the Laboratori Nazionali del Gran Sasso (LNGS) and has a 3.2~t liquid xenon inventory with a 2.0~t active target. Latest results from the XENON~1T experiment obtained with 34.2~days of exposure place a limit of $7.7 \times 10^{-47} {\rm cm}^{2}$ for a mass of  $35~{\rm GeV/c}^{2}$. Data was taken from September 2016 until it was halted after the October 30th 2016 central Italian earthquake. Data acquisition has since then resumed and Xenon purification has successfully reduced $^{85}$Kr contamination, so that the new dataset is now dominated by $^{222}$Rn backgrounds.
See figure~\ref{SI_limits} for the latest bounds XENON1T compared to other dark matter direct detection experiments. Note that since ICRC2017 PANDA-X presented new results at the TeVPA 2017 conference~\cite{PANDA-X_TeVPA} which improve upon the XENON limits.

In the future liquid nobel gas experiments plan to increase their target volume beyond the ton scale. The expected sensitivity with large volume Xenon TPC detectors is shown in figure~\ref{SI_limits} (middle) following the example of XENON~1T and XENONnT with 2~t and 6~t target mass, respectively. One potential difficulty with scaling up LXe experiments beyond the meter scale is that scintillation light at 175~nm wavelength is subject to Rayleigh scattering (scattering length $\sim 40$~cm). At ICRC the possibility to add fluorine, which may offer longer wavelength emission, while keeping the target radiopure was discussed~\cite{Kai_ICRC}. Longer wavelength scintillation light would be less attenuated and hence enable a reliable S1 signal time. Electron drift for charge extraction on the other hand will not be possible in the presence of fluorine and hence the method would only be applicable to single phase detectors.

XMASS is a single phase liquid xenon dark matter search experiment located at the Kamioka Observatory (overburden 2700 m.w.e) in Japan.
XMASS uses 642 high quantum efficiency PMTs (Hamamatsu R10789)  mounted in a spherical liquid xenon detector with radius of 40~cm. The detector is immersed in a 10~m diameter water tank that acts as an outer detector for active muon veto and a passive radiation shield against neutrons and gamma rays from the surrounding rock. XMASS combines a high light yield ($14.7$~pe/keV) with a low threshold (0.6~keVee) and is sensitive to $e/\gamma$-ray as well as nuclear recoil events. Data taking started in December 2010, but was interrupted from August 2012 till November 2013 for refurbishment of the detector. Preliminary limits obtained from the XMASS experiment on $\sigma^{\rm SI}$ are shown in figure~\ref{SI_limits} (right)~\cite{ICRC_XMASS_888}.


\subsection{Annual modulation}

\begin{figure}[htb]
        \centering
        \includegraphics[width=0.49\textwidth]{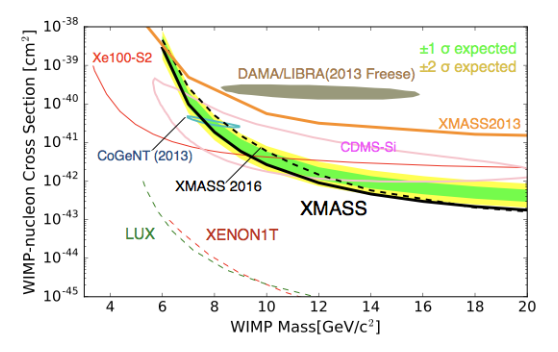}
        \includegraphics[width=0.49\textwidth]{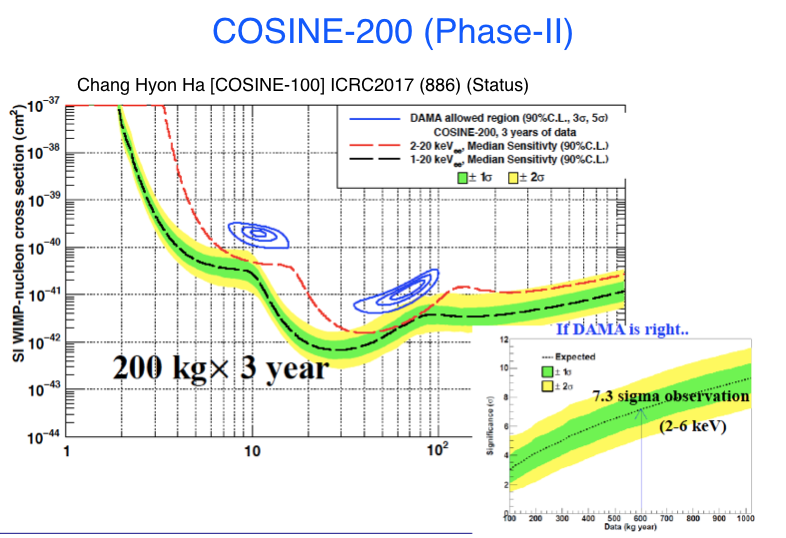}
        \caption{Left: Current limits obtained from the XMASS experiment in the search for annual modulations. Details are given in~\cite{ICRC_XMASS} (figure from~\cite{ICRC_XMASS}). Right: Expected sensitivity of the COSINE experiment~\cite{ICRC_COSINE}.
        }
        \label{annual_modulation}
\end{figure}

The event rate of a dark matter signal is expected to modulate annually due to relative motion of the Earth around the Sun. It would be a strong signature of dark matter. DAMA/LIBRA claims the observation of modulation at $9.3\sigma$ obtained with a total exposure of $1.33~{\rm ton}\times{\rm year}$ collected over 14 cycles~\cite{Bernabei:2013xsa}. The modulation amplitude is $0.0112 \pm 0.0012$~cpd/kg/keV for 2-6 keV. The observed annual modulation is inconsistent with indirect and other direct detection bounds, however different target materials have been used. Results of a search for an annual modulation in XMASS-I~\cite{Abe:2015eos,ICRC_XMASS} are shown in figure~\ref{annual_modulation} as an example.  Four efforts are now underway to search for the annual modulation using the same NaI target material as used by DAMA/LIBRA: (1) {\it ANAIS}~\cite{ANAIS} - Annual modulation with NAI Scintillators: Canfranc Underground Laboratory, Spain ; (2) {\it COSINE}~\cite{ICRC_COSINE}: Yangyang Underground Laboratory (Y2L), Korea ; (3) {\it PICO-LON}~\cite{PICOLON}  - Kamioka, Japan ; (4) {\it SaBRE}~\cite{SABRE} - Sodium Iodide with Active Background Rejection Experiment: Gran Sasso, Italy and Stawell Mine, Australia.

The COSINE experiment was extensively discussed at ICRC and has already started its physics data run. The COSINE experiment~\cite{ICRC_COSINE} is a joint venture between the DM-Ice~\cite{Cherwinka:2014xta} and KIMs collaborations~\cite{KIMS}. 
COSINE-100 consists of 8~NaI(Tl) crystals with a total mass of 106~kg, inside a 2000~L linear alkylbenzene (LAB)-based liquid scintillator (LS) veto, copper and lead shielding, and 37~plastic scintillator panels that serve as an additional cosmic-ray muons veto.
The detector is located at Y2L (see figure~\ref{Y2L}) and has been running smoothly since September 2016. Data-taking will continue for about two years before the next phase of the experiment (COSINE-200) will be assembled. The main goal of COSINE is to independently confirm or dispute the long-standing DAMA annual modulation signature. The latest sensitivity is shown in figure~\ref{Y2L}.


\section{Non WIMP Dark Matter}

\subsection{Searches for ALPs and Axions}

While WIMPs certainly remain the most explored option for explaining DM as extensions of the Standard Model of particle physics. In recent years interest in alternative scenarios involving weakly interacting sub-eV particles (WISPs), which could be non-thermally produced in the early universe and today be present as cold dark matter~\cite{Jaeckel:2010ni}, has grown. WISP describe particles such as axions, axion-like particles (ALPs), or hidden photons (HP). Axions are well-motivated as a solution to the strong CP problem. For a recent review on experimental searches see~\cite{Graham:2015ouw}.

\begin{figure}[htb]
        \centering
        \includegraphics[width=0.42\textwidth]{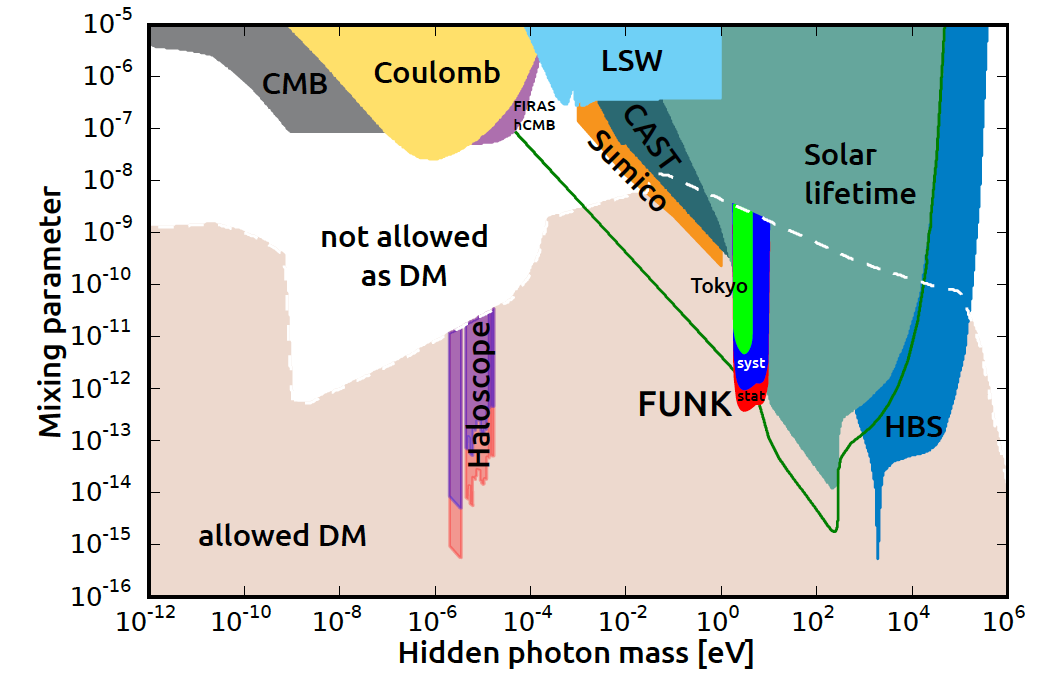}
        \includegraphics[width=0.57\textwidth]{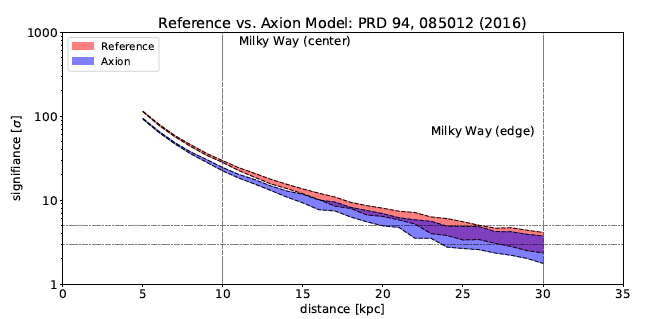}
        \caption{Left: The exclusion area in the mixing parameter vs. hidden photon mass parameter space are shown for the FUNK experiment in comparison to other results. FUNK results presented at ICRC are preliminary. Details on the figure can be found in~\cite{ICRC_FUNK}. Right: ALP sensitivity of IceCube with Galactic supernova. Details see~\cite{APL_IceCube} (Figures from~\cite{APL_IceCube,ICRC_FUNK}).}
        \label{FUNK_plot}
\end{figure}

Search for Axions and ALPs focuses on light-shining-through a wall (LSW) experiments, but they could also indirectly be detected for example through axion cooling in supernovae~\cite{Meyer:2016wrm}.
Axions can efficiently remove energy from a supernova explosion and hence contribute to the cooling of the system.
Neutrino production in a core collapse supernova (SN~II) is expected to be suppressed in the presence of axions, resulting modifications to the seconds-long tail of the SN~II neutrino light curve. A study of simulations of Galactic Type~II supernovae with and without the production of axions in the proto-neutron star following a reference model~\cite{Fischer:2016cyd} suggests that IceCube has a $5~\sigma$ discovery threshold within 3~kpc distance~\cite{APL_IceCube} as can be seen from figure~\ref{FUNK_plot}.

The FUNK experiment uses a 14~m$^2$ spherical mirror to search for hidden photons. HP are additional light U(1) gauge bosons that kinetically mix with the SM photons. Using a sensitive PMT at the radius point the experiment searches for faint electromagnetic waves that are emitted perpendicular to mirror surface when a HP crosses. Preliminary results are shown in figure~\ref{FUNK_plot} (details can be found in~\cite{ICRC_FUNK}).

\subsection{Nuclearites}

Nuggets of strange quark matter, aggregates of up, down and strange quarks in roughly equal proportions, could be produced in the first-order cosmological quark-hadron phase transitions in the early Universe or in processes related to compact stars such as neutron stars or quark stars~\cite{ICRC2017_924}. Neutral nuggets (covered by an electron cloud) are referred to as nuclearites.  Nuggets could be very massive and explain the observed dark matter in the Universe.  They are expected to have typical geocentric velocities of $\sim220$~km/s and might be detectable when they enter the earth atmosphere. Similar to meteors, nuclearites will produce visible light as they traverse the atmosphere, however as they are much more compact, the light along the track is very different and also the typical interaction height of 100~km and 10~km respectively. Currently a high sensitivity stereoscopic CMOS camera system is being tested to search for strange quark matter and interstellar meteorites. Interstellar meteoroids have geocentric velocities above 72~km/s compared to solar system meteoroids. The cameras are separated by a distance of 20.3~km, which allows to determine the altitude of the trace in the sky. The amount of light along the track also differs between the nuclearites and meteors. 
The system composed of 20 cameras with one year of data is expected to be
sensitive to nuclearite masses above about $10^2$~g at a flux sensitivity
of about $10^{-20}$ cm$^{-2}$ s${^-}1$ sr${^-1}$.
In the future space based observations by JEM-EUSO and EUSO-TA can boost sensitivities~\cite{JEM_EUSO_Nucleorites}.


\section{Conclusions}

The identity of dark matter remains one of the most outstanding mysteries in modern physics. All our knowledge about dark matter originates from astrophysical observations and cosmology, however to understand its nature a connection to particle physics needs to be found. 
Intense efforts to find dark matter are on-going with direct, indirect, and collider based searches. Dark matter could come in many different forms and shapes making discovery  prospects difficult to access. One of the main beacons from particle theory for the last decades, supersymmetry, has faded in light of tight LHC constraints. Dark matter searches have expanded in their diversity and are becoming a driving force behind many observatories.

At this ICRC conference no new smoking gun dark matter signal has emerged and past anomalies, such as the rise in the cosmic ray positron fraction observed by PAMELA and later precisely measured by AMS-02, the Galactic center gamma-ray excess, the 3.5~keV X-ray line, are now widely believed to be of non-dark matter origins. This progress, shows the healthy interaction of cosmic-ray and dark matter fields. Cosmic ray data is of high importance to the dark matter field as signals could hide in already existing data or relate to anomalies in direct searches.  
Non detections of dark matter reduce the available dark matter parameter space to inform future search strategies and help model builders to provide better guidance where to look next. 
Indirect dark matter search experiments are largely multi-purpose experiments that often combine guaranteed observational measurements with a discovery potential for dark matter. Only through an eventual dark matter signal will we be able to understand our Universe on a deeper level.


\section{Acknowledgements}
I would like to thank the organizers for the opportunity and honor to present the dark matter rapporteur talk. I am grateful for all the dark matter contributions to ICRC and like to thank the presenters for providing feedback and materials related to their contributions. I would like to thank Kev Abazajian, Saptashwa Bhattacharyya, Mathieu Boudaud, Marco Cirelli, Iris Gebauer, Sergio Hernandez, Fumiyoshi Kajino, Ranjan Laha, Hyunsu Lee, Kai Martens, Masayuki Nakahata, and Matthew Wood for materials and comments. I acknowledge support from the National Research Foundation of Korea~(NRF) funded by the Korea government (MSIP) Basic Science Research Program (NRF-2016R1D1A1B03931688).

\end{document}